\documentclass[a4paper, 11pt]{article}
\usepackage[left=2cm,top=1.9cm, bottom=1in, right=2cm]{geometry}

\usepackage{graphicx}
\usepackage{color}
\usepackage{algorithmic}
\usepackage{cite}
\usepackage{algorithm}
\usepackage{clrscode}
\usepackage{amsmath, amssymb, mathrsfs, amsfonts, amsthm}
\usepackage{epsfig, epstopdf}
\usepackage[tight,footnotesize]{subfigure}
\usepackage{stfloats}

\theoremstyle{theorem}
\newtheorem{theorem}{Theorem}%[section]

\newtheorem{definition}{Definition}

\newtheorem{remark}{Remark}

\theoremstyle{remark}

\def\msP{ \mathsf{P}}
\def\msT{ \mathsf{T}}
\def\mP{\mathbb{P}}
\def\sMLDA{\mathsf{MLDA}}
\def\sGA{\mathsf{GA}}

\DeclareMathOperator*{\argmin}{argmin}

\begin{document}

\title{Receivers for Diffusion-Based Molecular Communication:  Exploiting Memory and Sampling Rate}
\author{Reza~Mosayebi$^*$,
        Hamidreza~Arjmandi$^*$, Amin~Gohari$^*$, 
\\
        Masoumeh~Nasiri~Kenari$^*$ and  Urbashi~Mitra$^{\dagger}$\\
$^*$ Sharif Univ. of Tech. $^{\dagger}$ Univ. of Southern California (USC)
\iffalse
\thanks{The first four authors are with Electrical Eng. Dep., Sharif University of Technology, Tehran, Iran. emails: \{mosayebi,arjmandi\}@ee.sharif.edu, \{aminzadeh,mnasiri\}@sharif.edu. U. Mitra is with the Department of Electrical Engineering, University of Southern California (USC), Los angeles, 90089, USA e-mail: ubli@usc.edu 
}\fi
}

%\author{Sajjad~Beygi~\IEEEmembership{Student Member, IEEE}, Urbashi~Mitra~\IEEEmembership{Fellow, IEEE}\thanks{This research has been funded in part by ONR N00014-09-1-0700, NSF CNS-0832186 and NSF CCF-1117896. S. Beygi and U. Mitra are with the Department of Electrical Engineering, University of Southern California (USC), Los angeles, 90089, USA e-mail: \{beygihar,ubli\}@usc.edu .}}

% The paper headers
%\markboth{Submitted to ....}%

\maketitle

\begin{abstract}
In this paper,  a diffusion-based molecular communication channel between two nano-machines is considered. The effect of the amount of memory on performance is characterized, and a simple memory-limited decoder is proposed and
 its performance is shown to be close to that of the best possible imaginable decoder (without any restrictions on the computational complexity or its functional form), using Genie-aided upper bounds.  This effect is specialized for the case of  Molecular Concentration
Shift Keying; it is shown that a four-bits memory achieves nearly the same performance as infinite memory.
Then a general class of threshold decoders is considered and shown not to be optimal for a Poisson channel with memory, unless SNR is higher than a value specified in the paper.
{Another contribution is to show that receiver sampling at a rate higher than the transmission rate, i.e., a multi-read system, can significantly improve the performance.} The associated decision rule for this system is shown to be a weighted sum of the samples during each symbol interval. The performance of the system is analyzed using the saddle point approximation. The best performance gains are achieved for an oversampling factor of three.

\end{abstract}

\section{Introduction}
Successes in the design and development of nano-machines motivate the study of communication methodologies between such units. Interconnected nano-networks have potential applications in biomedical, industrial and environmentally engineered systems \cite{AAA, Nakano}.  Herein, we examine strategies for molecular communication, inspired by biological systems.  In particular, we examine a point-to-point communication system via molecular diffusion \cite{B5} (see \cite{New_2} for the motivation of using this type of communication).  A nano-transmitter releases molecules into the medium which are collected at a receiver.  Information can be transmitted by manipulating the number, type, and timing of the molecules.  Motivated by classical radio communication strategies, several molecular modulations have been recently introduced, including Concentration Shift Keying (CSK) (\emph{number} of molecules), Molecular Shift Keying (MoSK) (molecule \emph{types}), Molecular Concentration Shift Keying (MCSK) (a combination of the previous two) \cite{AA2, AA1}, the timing modulation (e.g. see \cite{Eckford22}) and  ratio-based modulations using properties of isomers \cite{Isomer}.  They are appealing because they use simple transmitter and receivers. 

Due to properties of diffusion, inter-symbol interference can occur (molecules from previous transmissions hitting the receiver after a long delay). By alternating molecule types, MCSK has zero interference from the previous symbol at the expense of two molecule types, but still suffers from interference from the other past symbols. This interference becomes significant at high data rates, corresponding to small symbol transmission periods.
 Performance degradation due to interference is studied in  \cite{B1,B2,B3}.
 The optimal receiver for a linear time-invariant propagation model is determined in \cite{AA3}.
 Unfortunately, these receivers demand high computational and memory complexities, making them  impractical for resource limited nano-machines.
%{\bf I'm confused about the refernce to \cite{MY1} as this is a capacity paper and not a MLSD paper --perhaps a different paper was intended?}
Maximum likelihood sequence detection is considerd in \cite{NewlyAdded} as well as sub-optimal and practical detectors.

\begin{figure}
\begin{center}
\includegraphics[width=167mm]{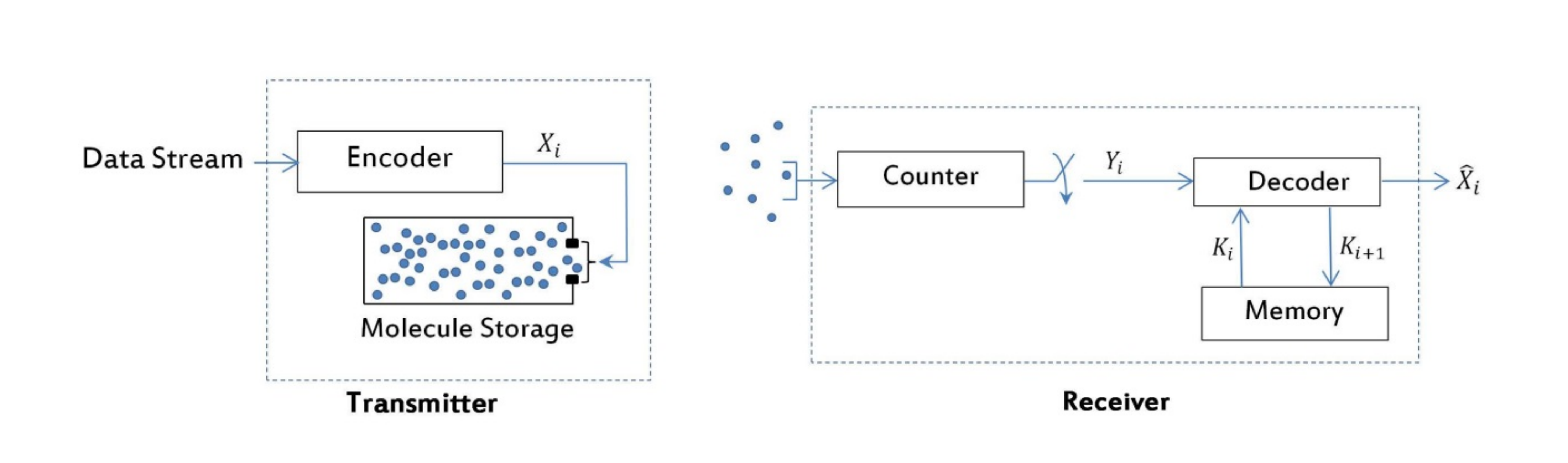}%{SchematicReceiverTransmitter.jpg} %.pdf
\end{center}
\vspace{-1cm}
\caption{Block diagram of a transmitter with gate controlled molecule storage, and a general slotted receiver}
\label{fig:transmitterschematic}
\vspace{-0.5cm}
\end{figure}

In this paper we consider the communication model introduced in \cite{AA1} and depicted in Fig. \ref{fig:transmitterschematic}, where the transmitter has a storage of molecules. The storage has an outlet whose size controls how many molecules can exit the storage and diffuse in the environment. The outlet of the storage can be controlled; it opens for a short period and the size of the opening in time slot $i$, $X_i$ is specified by the encoder for signaling. The released molecules diffuse through the environment (a diffusion channel) and a subset of the released molecules hits the receiver. The receptors at the receiver make chemical bonds with the received guest molecules and initiate a pre-defined process. The general block diagram of the receiver is depicted in Fig. \ref{fig:transmitterschematic}.  The front-end of the receiver is a counter, which tracks the number of received molecules.
 The decoder is allowed to read the counter and reset it. The decoder makes its decision about the transmitted symbols based on the read counts.

{Transceiver complexity is a significant issue for communication between nano-machines \cite{Re1}}.  As such, we restrict our attention to transceiver systems that have simple implementations, exploit limited memory, and as a result a limited decoding delay.  In particular, we assume that uncoded message symbols are transmitted, one-by-one.  The receivers have either zero or one symbol delay, with a 
  $B$ of bits of memory, which can store any  $B$ bit function of the previously decoded symbols. 

% More precisely by a zero-delay decoder %with $B$-bits of memory we mean %one that reconstructs the transmitted message %symbol at stage $i$, i.e. $\hat{M}_i$, based on the received signal at time %$i$, i.e. $Y_i$, and the $B$ bits that are stored in its memory (which is %iteratively updated and is a function of past observations $Y_{i-1}, Y_{i-2}, %\cdots $). 

To justify our assumptions on the transmitter and receiver, we note that 
\begin{enumerate}
\item Our assumptions form a restriction on the complexity of the transmitter and receivers thus rendering our proposals practical for implementation in real nano-machines.  In fact, the modulation schemes proposed to date (CSK, MoCSK, PPM, {etc.}) are of interest due to their associated simple and memorlyess decoders \cite{AA2,AA1}.

\item Limited decoder memory is equivalent to limited storage, which in turn, implies limited decoding delay to limit the loss of data due to storage overflow.  

\item While our assumptions are simplifying with regards to complexity, they do not limit our ability to focus on the unique nature of the diffusion channel.  In particular, we have a discrete channel with memory:  residual molecules from prior transmissions can be coincident at the receiver with the current molecular transmissions.  Since Brownian motion is used to model the diffusion channel, and the information is encoded in the \emph{average} diffusion rate, the overall channel is non-linear.  In fact, non-linear channels with memory are notorious for their lack of tractability.

% They allow us to focus on the nature of the memory of the diffusion channel. %The diffusion communication channel is a particular discrete channel with %memory, as molecules released from the transmitter can reach the decoder %after a long delay. Thus residual particles from previous transmissions, %as well as molecules that have been just released, may hit the receiver. %We note that if the channel was memoryless, evaluating the performance of %an uncoded transmission and zero-delay decoding schemes would have been %trivial. Unlike discrete memoryless channels whose analysis is tractable, %non-linear channels with memory are notorious for not being amenable to %tractable calculations.\footnote{While the diffusion channel is linear, %the overall channel from the size of the outlet to the number of molecules %received is not.} For instance 

\item  We underscore that our assumption of $B$ bits of memory at the decoder is not as restrictive as one might think. We do not specify the form of the bits; that is, the decoder can store {\bf any} $B$ bits of prior information.  The only restriction is that the memory has to be updated sequentially and causally as the receiver obtains a new output symbol. One of our objectives is to determine what should be stored and understand performance gains as a function of increasing memory.

\end{enumerate}
{
To the best of our knowledge, this work is the first one to study the fundamental limits of communication with memory-limited devices (previously limitations on power, delay, etc have been addressed in the literature of information theory or communication theory).}
In this work, we make four key contributions for the  {two main building blocks of the  receiver depicted in Fig. \ref{fig:transmitterschematic}}.  \emph{For the decoder block:} (1) We propose the memory-limited decision aided (MLDA) decoder which exploits $B$ bits of memory; (2) we show the near-optimality of the MLDA decoder by comparing its performance with the performance of a Genie-aided optimal decoder (without any restriction on the computational complexity
or functional form of the stored $B$ bits).
  In providing these results, we
also study how information (or the lack thereof) about the interference affects
decoder performance; (3) {we prove analytical results, some of which provide surprising insights into this problem: for instance, we show that a natural threshold decoder is not optimal for a Poisson channel with memory. On the other hand, we show in Theorem 2 that ML decoder is a threshold decoder for SNRs above a threshold (which we find an analytical lower bound on). }
 \emph{For the counter block:} (4)  {we propose the use of multiple samples during each symbol period (multi-read system)} and show that this oversampling significantly decreases the probability of error. In contrast, most previous works had assumed that the counter is read only once every transmission period. We first show that the associated MLDA test statistic is a weighted sum of the samples read within a symbol interval.    We also analytically evaluate the performance of the multi-read receiver using a saddle point approximation as the test statistic distribution does not admit a closed form.  The accuracy of the approximation is verified via simulation.

The rest of paper is organized as follows: In Section \ref{sec:com-med}, a diffusion-based molecular channel is defined. Section \ref{definition:mlzd} defines the class of transmitter and receivers that we are interested to study. In Section \ref{sec:simpleReciever}, a particular simple and practical decoder is introduced and its probability of error is analysed. In Section \ref{sec:MRMLD}, the main results of this paper on memory-limited decoders are provided, with the proofs given in Appendices A and B. Section \ref{sec:CB} makes a case for having multiple reads in the counter block of a receiver.  Section \ref{sec:MCSK} briefly reviews and comments on the MCSK modulation scheme. Application of our main results to MCSK is given in the same section. Finally Section \ref{sec:conclusion} concludes the paper.

\section{System Model}\label{sec:com-med}
\subsection{Transmitter and channel models}
{
We consider an arbitrary diffusion-based molecular communication medium, uniform or non-uniform, with drift or without drift, and with physical barriers or in an open space. The main assumption that we make about environment is that it is stationary, meaning that the diffusion coefficients in each location do not change over time.}
 Due to extensive molecule propagation times between transmitter and receiver, the effective channel has {\em memory} resulting in interference, which will be more fully discussed in the sequel.  

We assume that the transmitter has a storage of molecules. The storage  has
an outlet whose size controls how many molecules can exit the storage to
be diffused in the environment. Since the number of molecules is large and
the probability of each exiting the storage is small, the number of molecules
that exit from the storage in a given time slot follows a Poisson distribution.
This is the Poisson model proposed in \cite{AA1} (see \cite{New_2} for a Poisson model of
the environmental noise) and we adopt it in this
work. We use $\mbox{Poisson}(X)$ to denote the number of molecules
exiting the storage (and diffused in the environment). The
parameter $X$ is called the diffusion rate and is determined
by the size of the opening of the outlet. Parameter $X$ itself is
a random variable that is selected by the transmitter based on
the message it wishes to transmit. Thus, we have a doubly stochastic Poisson process;
the rate of the Poisson is the information signal, which in itself is random due to our models on the information sequence. 

We assume that the transmission occurs in  time slots of equal duration; this is called the {\em symbol
duration} and is denoted by $t_s$. We assume that the transmitter may only set
the size of the storage's outlet at the beginning of time slots of length $t_s$.
We assume that the transmitter chooses a diffusion rate  of $X_i$ at time slot
$i$.
Assume that a molecule is released from storage at the beginning of a time slot, and is absorbed the first time it hits the receiver. 
The probability that the molecule hits the receiver during the initial time slot of duration $t_s$  is denoted by $p_1$; more generally, the probability of hitting during the next $k$-th time slot
is denoted by $p_k$, $k\geq 1$. {The values of $p_k$'s depend on the communication medium, which can be in either 1-d or 3-d.} Note that we assume that the receiver and the transmitter are synchronous.

Generally, the signal is decoded at the receiver based on the
total number of molecules received during the individual time slots.
We assume that in a time slot, three sources contribute to the
received molecules: (i) molecules due to
the transmission in the current time slot, (ii) the residue molecules
due to the transmission in the earlier time slots known as the interference
signal, and (iii)  noise molecules, described below.
We assume that inter-molecule collisions
have little effect on the molecule's movement. 

Within our reception time slot of duration $t_s$,
there is noise due to molecules in the environment that are
not part of our transmission but of the same type as  the transmitted molecules.  We
model the number of noise molecules within a slot of duration $t_s$ as  a Poisson
random variable with parameter $\lambda_0 t_s$.

Assume that the transmitter selects a diffusion rate  of $X_i$ at time slot
$i$. 
Based on the thinning property of Poisson distribution, when we use one molecule
type in all time slots, the number of received molecules in the time slot
$i$, denoted by $Y_i$, can be written as follows:
\begin{align}
Y_i&\sim \mbox{Poisson}( \lambda_0 t_s + \sum_{k=0}^\infty p_{k+1} X_{i-k}  )= \mbox{Poisson}(p_1X_i+I_i), i\geq 1 \label{eq:Y_1}
\end{align}\normalsize
where $I_i= \lambda_0 t_s + \sum_{k=1}^\infty p_{k+1} X_{i-k}$ is the sum of interference and noise
at time $i$.  {Further, we assume $X_i=0$ for $i\leq 0$ (i.e. transmission starts at time $i=1$), hence summation is from $k=0$ to $\infty$.}

One of our innovations over our prior work in \cite{AA1} is that we explicitly consider more interference terms, {\em i.e.} in \cite{AA1}, we argued that a large $t_s$ results in negligible $p_k$ for $k>2$ and thus could ignore interference beyond the previous time slot.  A further innovation herein is the consideration of multiple reads (sampling within a single time slot) which also results in improved performance.

\vspace{-0.5cm}
\subsection{Decoder Framework}
We consider a simple, practical decoder structure for molecular communication. To this end, we make two key assumptions about memory and delay.  The first is that the transmitter transmits molecules at rates $X_i$ independently and thus there is no error correction coding at the transmitter.  Most generally, we can assume that $X_i \in \left \{ s_0, s_1, \cdots, s_{L-1} \right\}$.  Such $L$-ary signaling implies a bit rate of $\frac{\log{(L)}}{t_s}$.  For clarity of exposition we assume $L=2$ and that $X_i = 0$ or $X_i = s$ with probability $1/2$.

To frame our decoder, we observe that in the case of no interference, the maximum a posteriori (equivalently maximum likelihood due to the equal priors) receiver is given as follows. Our observation $Y$
 is distributed as $\mbox{Poisson}(\lambda_0 t_s +p_1X)$, where $X \in \{0,s\}$ and  {the associated probability mass functions (pmfs)} are
\begin{align}
\mathbb{P}(Y=y|X=x)&=\frac{e^{-(p_1x + \lambda_0 t_s) }(p_1x+ \lambda_0
t_s)^y}{y!}, \qquad x\in \{0,s\}.
\end{align}
The optimal detector is of a threshold form where if $y \leq \tau$ we select
$\hat{X} = 0$ (and $\hat{X}=1$ if the observation exceeds the threshold),  where
$
\tau= \frac {p_1 s}{\ln(1+\frac{p_1s}{\lambda_0 t_s})}.
$

In the presence of interference, the optimal decoder cannot be reduced to a simple threshold structure. However, threshold detectors are simple to implement.

\emph{Memory-Limited, Zero-Delay Decoders:}\label{definition:mlzd}
In the sequel, we introduce a simple decoder design based on the storage of previous decisions.  In order to assess the performance of this decoder, we need to ask what can be accomplished with $B$ bits of side information.  To formalize this notion, we have the following:

A \emph{memory-limited} decoder, with $B$ bits of memory, can store a number
in the set $\{1,2,\cdots, 2^B\}$ (or $[1:2^B]$ as a shorthand)  for future use. We use random variable $K_i$ to denote
the value stored in the memory at time slot $i$ for $i\geq 1$. We assume
that initially at time slot $0$, $K_0=0$. At time instance $i$, the decoder
uses the received signal $Y_i$ and the memory register $K_i$ to update the
memory register $K_{i+1}$, i.e. $K_{i+1}$ is constructed as a function (or
stochastic function) of $K_i$ and $Y_i$. In other words, we use an ``update"
function $f_i:[1:2^B]\times\mathbb{Z}_{+}\mapsto [1:2^B]$ and construct
$K_{i+1}=f_i(K_i, Y_i)$. Here $\mathbb{Z}_{+}$ is the set of non-negative
integers. Observe that the updating function $f_i$ may be time-varying: for
example, for even values of $i$ $f_i$ may be one function and for odd values
it may be another function.

A decoder is called \emph{zero-delay}, if it computes the reconstruction
$\hat{X}_i$ at the end of time slot $i$. 
A memory-limited zero-delay decoder computes the reconstruction $\hat{X}_i$
as a function of $K_i$ and $Y_i$ only, i.e. $\hat{X}_i=g_i(K_i, Y_i)$ for
some function $g_i:[1:2^B]\times\mathbb{Z}_{+}\mapsto \{0, 1\}$. A decoder
is called a \emph{zero-delay threshold} decoder, if the functions $\{g_i\}_{i=1:\infty}$
have a threshold rule: for each value of $i$ and $K_i=k$, we compare $Y_i$
with some threshold $\tau_{ki}$ and set $\hat{X}_i=0$ if
$Y_i\leq \tau_{ki}$ and set $\hat{X}_i=s$  if $Y_i>\tau_{ki}$.

A decision rule for a  \emph{memory-limited} and \emph{zero-delay} decoder
is specified by a sequence of update functions $f_i:[1:2^B]\times\mathbb{Z}_{+}\mapsto
[1:2^B]$, and a sequence of decoding functions $g_i:[1:2^B]\times\mathbb{Z}_{+}\mapsto
\{0,1\}$.

Given  a memory of size $B$, we are interested in finding the optimal memory-limited
decoder, and see how its performance behaves with $B$. We intuitively expect
the memory to contain information about the interference, useful for decoding
the current symbol. More precisely, if we denote the { interference plus noise} at time
slot $i$ by
$
I_{i}= \lambda_0t_s+\sum_{j=1}^{\infty} p_{j+1} X_{i-j} 
$
then given $K_i=k$, one can compute the posterior probability distribution
of $p(I_i|K_i=k)$. The induced $p(I_i|K_i=k)$ and $Y_i$ turn out to be
sufficient statistics for reconstructing $\hat{X}_i$.\footnote{This is because
the MAP decision rule compares $p(Y_i=y|X_i=x, K_i=k)$ for different values
of $x$. But this probability is equal to 
$
\mathbb{E}_{I|K_i=k}\frac{e^{-(p_1x+I)}(p_1x+I)^y}{y!}
$
which can be calculated using the conditional pmf of $p(I_i|K_i=k)$.
}

\begin{definition}
The minimal probability of error of a \emph{memory-limited} and \emph{zero-delay}
receiver with rate $R$, transmission amplitude $s$ and memory size $B$, $\msP
(R, s, B)$, is defined as the minimum of the probability of error
of all decoders with memory $B$ (specified by  update functions $f_i$ and
a decoding functions $g_i$). The minimal probability of error of a \emph{memory-limited}
and \emph{zero-delay threshold} receiver with rate $R$ and memory size $B$,
$\msP_\msT(R, s, B)$, is defined similarly where we restrict 
to \emph{zero-delay threshold} decoders. 
\end{definition}
\begin{remark} By definition $\msP_{\msT}(R,s, B)\geq \msP (R, s,
B)$. Clearly both $\msP (R, s, B)$ and $\msP_{\msT}(R, s,
B)$ are decreasing functions of $B$, i.e. the more memory, the less probability
of error.%How does it behave with $s$ and $R$?
\end{remark}

\vspace{-0.7cm}
\section{Proposed receiver: Decoder}\label{sec:simpleReciever}
In this section, we first provide a simple and practical decoder and then show how performance for the proposed decoder can be easily computed. In order to compare this performance with $\msP
(R, s, B)$ and $\msP_T
(R, s, B)$, we will later examine a genie-aided decoder.
\subsection{Memory Limited Decision Aided Decoder (MLDA)}
 Consider the simple strategy of
storing the decoded rates $(\hat{X}_{i-1}, \hat{X}_{i-2},
\cdots, \hat{X}_{i-B})$
in  memory. These $B$ bits are used to estimate the interference rate
 which is equal to $ p_2X_{i-1}+p_3X_{i-2}+\cdots+p_{B+1}X_{i-B}+\sum_{j=B+1}^{\infty}
p_{j+1} X_{i-j}.$
The updating function $f_i$ for this decoder is time-invariant, i.e. $f_i=f$,
and can be easily implemented using a shift register. 

It remains to specify the decoding function $g_i$ which shall also be time-invariant
 i.e. $g_i=g$. To derive the decoder, let us assume
for a moment that the decoder knows the exact value of interference, i.e.
it knows that $I_i=\alpha$. To recover $X_i\in\{s_0=0, s_1=s\}$ we
use a MAP decoder and compute
\begin{align}
\max_{j} {\mathbb{P}(Y_i=y|X_i=s_j, I_i=\alpha)}  
= \max_{j} \left(\frac{e^{-(p_1s_j + \alpha) }(p_1s_j+  \alpha)^y}{y!}\right).
\end{align}
Observe that the above equation is an ML decoder rule, which is equivalent to MAP because $X_i$ is $\mathsf{B}(\frac 12)$ and independent of $I_i$ (a function of $X_{i-1}, X_{i-2}, \cdots$).

The resulting rule is a threshold rule: we set $\hat{X}_i=0$ if $y\leq \tau$, and set $\hat{X}_i=s$ if $y>
\tau$ where
\begin{align}\label{eqn:eqnTh}
\tau=\frac{p_1 (s_1-s_0)}{\ln(\frac{p_1s_1+ \alpha +\lambda_0 t_s}{p_1s_0+
\alpha+ \lambda_0 t_s})}
\end{align}

As we do not know $I_i = \alpha$ exactly, we approximate $I_i$ given our previously decoded rates, that is,
\begin{align*}
\hat{I}_i&=\lambda_0t_s+p_2\hat{X}_{i-1}+\cdots+p_{B+1}\hat{X}_{i-B}+\sum_{j=B+1}^{\infty}
p_{j+1} \mathbb{E}[X_{i-j}]
\\&=\lambda_0t_s+p_2\hat{X}_{i-1}+\cdots+p_{B+1}\hat{X}_{i-B}+\frac {s_0+s_1}{2}\sum_{j=B+1}^{\infty}
p_{j+1}.
\end{align*}

In other words, we use the threshold given in equation \eqref{eqn:eqnTh},
when $\alpha$ is replaced by the expression given above. Clearly the value
of the threshold would depend on $(\hat{X}_{i-1}, \hat{X}_{i-2}, \cdots,
\hat{X}_{i-B})$ that are stored in the memory.

This simple decoder is of low complexity for implementation on a nano-machine.
Note that the decoding rule described above is not necessarily the optimal
MAP decoding rule,  {but numerical results show this simple decision rule performs quite close to the optimal one.}

\subsection{Computation of the MLDA Probability of Error}
\label{subsubsection:DPE}

We find an analytical expression for the probability of error of the simple
decoder. We have
\begin{align}
P_e=\frac 12\big(\mathbb{P}(E|X_i=s_0)+\mathbb{P}(E|X_i=s_1)\big)=\frac 12\big(\mathbb{P}(Y_i>\tau
|X_i=s_0)+\mathbb{P}(Y_i\leq \tau |X_i=s_1)\big),
\end{align}
where $E$ is the error event. As the previously decoded symbols are utilized
in the current time slot, false decoding in the previous time slots may propagate
to decoding in the current time slot. Therefore, it is necessary to consider
error propagation in computing the probability of error. 

We compute the probability of error for the special case of $B=1$, i.e. storing
only the last decoded symbol. Analytical expressions for the probability
of error for larger values of $B$ can be derived similarly. For $B=1$, the
estimated value of interference plus noise is equal to $$\hat{I}_i( \hat{X}_{i-1})=\lambda_0t_s+p_2\hat{X}_{i-1}+\frac{s_0+s_1}{2}\sum_{j=2}^{\infty}
p_{j+1}=\lambda_0t_s+p_2\hat{X}_{i-1}+\frac{s_0+s_1}{2} (1-p_{1}).$$ We have
the following equation for $k=0,1$:
\begin{align}
\mathbb{P}(E|X_i=s_k) 
&=\sum_{x_{i-1},\hat{x}_{i-1}}{p_{X_{i-1},\hat{X}_{i-1}|X_i}(x_{i-1},\hat{x}_{i-1}|X_i=s_k)\mathbb{P}(E|X_i=s_k,x_{i-1},\hat{x}_{i-1})}\nonumber\\
&=\sum_{x_{i-1},\hat{x}_{i-1}}{p_{X_{i-1}}(x_{i-1})p_{\hat{X}_{i-1}|X_{i-1}}(\hat{x}_{i-1}|x_{i-1})\mathbb{P}(E|X_i=s_k,x_{i-1},\hat{x}_{i-1})}.\label{eq:aa1}
\end{align}
Note that 
\begin{align}
\mathbb{P}(E|X_i=s_1,x_{i-1},\hat{x}_{i-1})&=\sum_{\alpha,~ y\leq \tau} \mathbb{P}[I_i=\alpha|x_{i-1}]\frac{e^{-(p_1s_1
+ \alpha) }(p_1s_1+ \alpha)^y}{y!}
\\
\mathbb{P}(E|X_i=s_0,x_{i-1},\hat{x}_{i-1})&=\sum_{\alpha,~ y>\tau} \mathbb{P}[I_i=\alpha|x_{i-1}]
\frac{e^{-(p_1s_0 +  \alpha) }(p_1s_0+ \alpha)^y}{y!}
\end{align}
in which $\tau=\frac{p_1 (s_1-s_0)}{\ln(\frac{p_1s_1+ \hat{I}_i( \hat{x}_{i-1})}{p_1s_0+
\hat{I}_i( \hat{x}_{i-1})})}$. The conditional pmf $\mathbb{P}[I_i=\alpha|x_{i-1}]$
can be computed as follows:
$$\mathbb{P}[I_i=\alpha|x_{i-1}]=\mathbb{P}\left[\sum_{j=2}^{\infty} p_{j+1}
X_{i-j}=\alpha-\lambda_0 t_s -p_2x_{i-1}\right].$$
Therefore, $\mathbb{P}(E|X_i=s_0)$ and $\mathbb{P}(E|X_i=s_1)$ can be computed
recursively from (\ref{eq:aa1}) which is a system of linear equations.

\section{Performance Bounds}\label{sec:MRMLD}
As in previous sections, we focus on the binary case $X_i \in \{s_0=0,s_1=s\}$.  Ideally we wish to make concrete statements on optimized general and threshold decoders, that is by optimizing over all $g_i$ and $f_i$.  It is clear that the probability of error of such optimized decoders is upper bounded by that of the MLDA introduced in Section \ref{sec:simpleReciever},  i.e. if we denote
the MLDA performance by $\mathsf{MLDA}(R, s, B)$, we have that $\sMLDA(R, s, B)\geq \msP_\msT(R,s, B)\geq  \msP(R,s, B)$.
\color{black}
 As a lower bound to $\msP
(R,s, B)$ we introduce a genie-aided decoder which knows the interference exactly, this results in the threshold rule provided in eqn. \eqref{eqn:eqnTh} where $\alpha$ is the value of interference. We denote this genie-aided decoder by ``ML decoder with known interference" and show its performance by $\sGA(R,s)$.\color{black} As a more meaningful genie-aided decoder for determining a lower bound, we imagine (a threshold, or general)  decoder that has access to a genie who can pass any $B$ bits of information about the interference to the receiver. Since the $B$ bits are arbitrary, this new genie-aided decoder (that optimizes over the $B$ bits of side information) provides a tigher lower bound. \color{black}We need to examine how we can exploit $B$ bits of side information. Determining this genie-aided decoder is non-trivial. To appreciate the challenge of our evaluation, observe that to determine the side information by using brute force, one needs to consider the set of values that  the interference, i.e., 
$\sum_{j=1}^{\infty} p_{j+1} X_{i-j}$, can take. The genie
should specify a mapping from the set of values to the set $\{1,2,\cdots,
2^B\}$, {\em i.e.} for each value of interference it should specify the $B$ bits
that it is revealing to the decoder. To find the optimal performance that
a genie can achieve, one needs to optimize over all such mappings. This is
computationally cumbersome, if possible, even for small values of $B$.  Theorem
\ref{thm:thm3} below simplifies calculation of the genie-aided memory limited lower bound for the optimum threshold decoder, i.e., $\msP_\msT(R, s, B)$, and reveals an intuitive
structure:  the optimized threshold decoder with $B$ bits of side information optimally
quantizes the interference. 

The general case is more complex.  Consider an example where we can enumerate over all possibilities of the interference.  Let
 $s=400$ and the interference takes values uniformly from
the set $\mathcal{I}=\{\alpha_1, \alpha_2, \alpha_3, \alpha_4\}$ where $\alpha_1=1<\alpha_2=39.6899<\alpha_3=65.2742<\alpha_4=103.9640$, and the genie can provide one bit to the receiver. Then telling the receiver
whether interference is in the set $\{\alpha_1, \alpha_4\}$ or $\{\alpha_2,
\alpha_3\}$ is strictly better than telling the receiver if interference
is less than a number or not ( {\em i.e.} interference is high or low)! 
The next theorem shows that for threshold decoders, however, the phenomenon
discussed in the above observation does not happen.
 
\begin{theorem}\label{thm:thm3}
In a threshold decoder (comparing received signal $y$ with a threshold), the optimized genie-aided decoder with $B$ bits of side information  quantizes the interference values into $2^B$ bins and uses the interference interval's bin index as side information. 
\end{theorem}
\begin{remark}
The above theorem significantly reduces the computational complexity of finding the optimal Genie for computing the  lower bound on $\msP_\msT(R,S,B)$, denoted by
$\sGA_{\msT}(R, s, B)$. \color{black} It implies that one needs to only specify a set of $2^B-1$ points in $\mathbb{R}$
(hence dividing $\mathbb{R}$ into $2^B$ intervals), and read off the mapping
according to the index of the interval where interference falls into.  
\end{remark}

We discussed lower and upper bounds on $\msP (R, s, B)$ and $\msP_\msT(R,
s, B)$. In the following theorem we discuss a relation between these two.
\begin{theorem}\label{thm:thm2}
Given {any communication medium}, choose a sufficiently small rate $R$ such that hitting probabilities satisfy $p_1>(e-1)\sum_{j=2}^{\infty}p_j$. Then given any noise level $\lambda_0$, there exists some $s^*$ such that $\msP (R,s, B)=\msP_\msT(R, s, B)$
for all $s>s^*$ regardless of the value of $B$. More specifically, $\msP
(R,s, B)=\msP_\msT(R, s, B)$ holds if $s$ satisfies the following two inequalities
\begin{align*}
\frac{\lambda_0 + Rp_1s}{Rs} \ln\left(\frac{\lambda_0+ Rs\sum_{j=1}^{\infty}p_j}{\lambda_0+ Rs\sum_{j=2}^{\infty}p_j}\right)\geq
p_1 \geq \max\left[\frac{\sum_{j=1}^{\infty}p_j}{2}, \frac{\lambda_0 + Rs\sum_{j=2}^{\infty}p_j}{Rs} \ln\left(\frac{\lambda_0
+ Rp_1s}{\lambda_0}\right)\right].
\end{align*}
\end{theorem}
 The proofs for the above theorems are given in the Appendix.

\textbf{Numerical Results:} The above discussion indicates that $\sMLDA(R,s,B)\geq \msP_\msT(R, s, B)\geq \sGA_\msT(R,s,B)\geq \sGA(R,s)$, and $\sMLDA(R,s,B)\geq \msP(R, s, B)\geq \sGA(R,s)$. Since computing $\msP_\msT(R, s, B)$ and $\msP(R, s, B)$ are computationally intractable, herein  we compare
 the performance of the $\sMLDA(R,s,B)$  with the two  genie-aided decoders, i.e. with $\sGA_{\msT}(R,s,B)$ and $\sGA(R,s)$, showing that they come close to each other and effectively sandwitching the desired $\msP_\msT(R, s, B)$ and $\msP(R, s, B)$.\color{black} {Here and in the following numerical results, except for Fig. \ref{Fig3d}, we use a one-dimensional  Brownian motion without drift, as in \cite{New_2,Eckford22}, since for this motion, closed form expressions for hitting probabilities $p_k$'s can be found.}
 For a one-dimensional  motion without drift, the hitting probabilities can be found as follows\cite[5.3]{Book}:
$p_1=2Q\left(\frac{\rho}{\sqrt{t_s}}\right)$, and 
\begin{align}p_k=2Q\left(\frac{\rho}{\sqrt{kt_s}}\right)-2Q\left(\frac{\rho}{\sqrt{(k-1)t_s}}\right),
\quad k=2,3,\cdots.\label{eqn:AA1}\end{align}
where  $\rho \triangleq d/\sqrt{2D}$  is a single parameter that summarizes
how the diffusion constant ($D$) and the distance between the transmitter and
the receiver ($d$) affect the communication channel between the transmitter and
the receiver.  The hitting probabilities
depend on $R$ and $\rho$ ( {representing transmission rate and channel parameter, respectively}) only through $\rho^2R=\rho^2/t_s$. So, we can set a particular value for $\rho$
and discuss the gap between the lower and upper bounds in terms of the rate
$R$. A practical
value for $\rho^2$ is $0.3$ \cite{Web}. 

Figure \ref{fig:CSK2} (a) depicts $\sMLDA(R,s,B)$ and  $\sGA_{\msT}(R,s,B)$ for $B=1,2$ and $\sGA(R,s)$
 in terms
of $R$ for  $s=60$ and noise power $\lambda_0=0.4$. The subfigure (a) is for CSK modulation, which is the setup we have used so far. The top two curves are for $B=1$, the
two bottom curves are for $B=2$, and the lowest curve is for $\sGA(R,s)$ (known interference).  The range of $R$ is chosen such
that the error probabilities are in a reasonable range. This shows that the
lower and upper bounds are close to each other. It can be seen that  $\sMLDA(R,s,B=2)$
is lower than $\sGA_{\msT}(R,s,B=1)$. The curves become much closer to each other
(depicted in the Fig. \ref{fig:CSK2} (b)) when we apply the lower and upper bounds to
the MCSK modulation scheme  as will be further discussed in Section \ref{sec:MCSK}.   {To generate the results of this figure, the threshold values for computing  $\sGA_{\msT}(R,s,B)$ have been computed using a brute-force search.}

\begin{remark}
In Fig. \ref{fig:CSK2}, the condition of Theorem \ref{thm:thm2}
is satisfied. Therefore the lower bound curves for the threshold decoder ($\sGA_\msT(R,s,B)$) are
also lower bounds for $\msP (R, s, B)$.
\end{remark}
\iffalse
\begin{figure}
\begin{center}
\includegraphics[width=180mm]{All_csk_2}      %%%%%%%%%%%%%%%%%%%% 6
\includegraphics[width=180mm]{All_2}      %%%%%%%%%%%%%%%%%%%% 6
\end{center}
\vspace{-1.0cm}
\caption{\footnotesize{(a): Main Model (CSK), (b): MCSK. Lower and upper bounds for limited memory receivers.
\iffalse The curves labeled as ``Threshold Dec." are lower bounds on $E_T(R, s, B)$, and ones labeled as ``Known Int." are the lower bounds on $E(R, s, B)$.  Parameters for this figure: $s=60$ and $\lambda_0 =0.4$
Not revised\color{black}\fi
}
}
\label{fig:CSK2}
\vspace{-0.5cm}
\end{figure}

\fi
\begin{figure}
\centering
\subfigure[]{
 \centering
\includegraphics[width=0.9\linewidth]{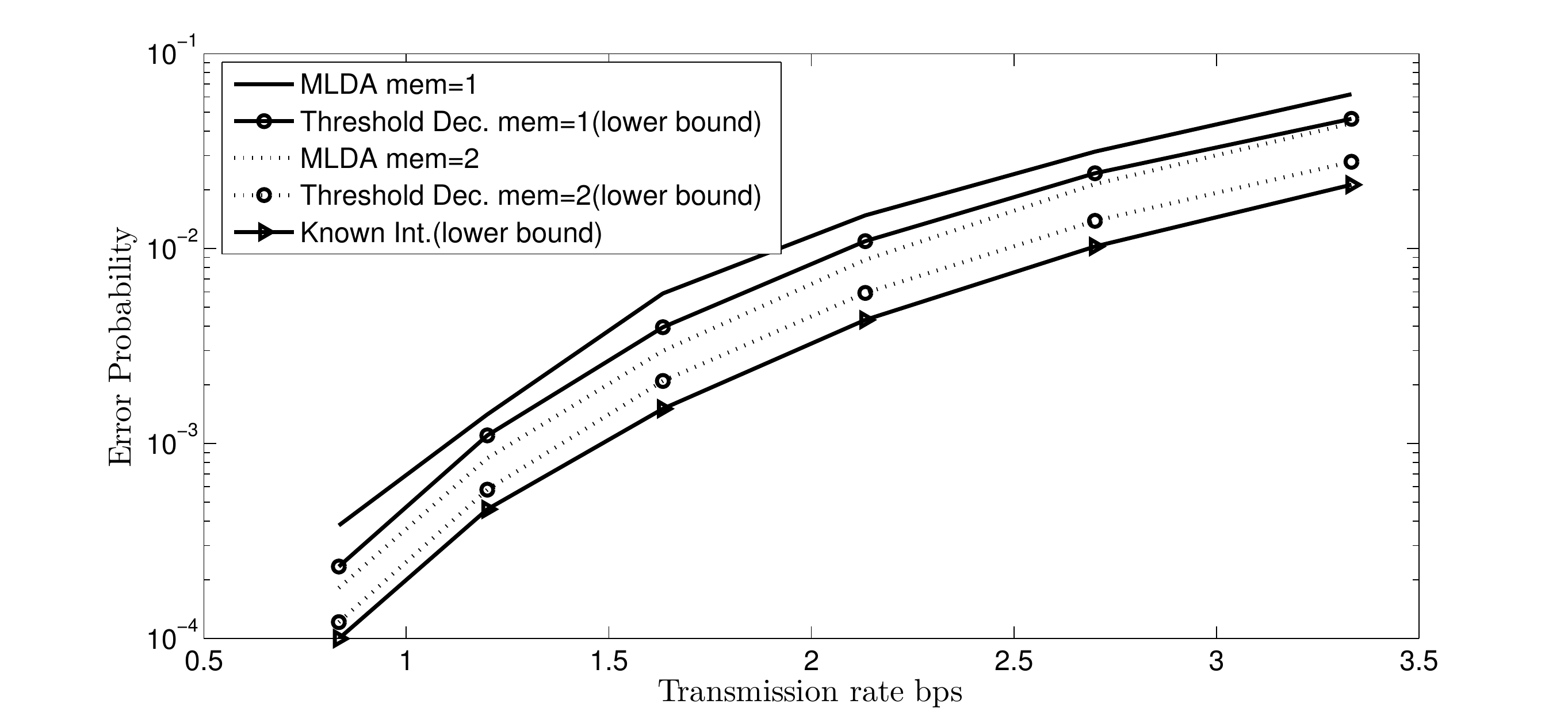}      %%%%%%%%%%%%%%%%%%%% 6 
\label{fig:all_csk_2}
}
\vspace{-0.4cm}

\subfigure[]{
\centering
\includegraphics[width=0.9\linewidth]{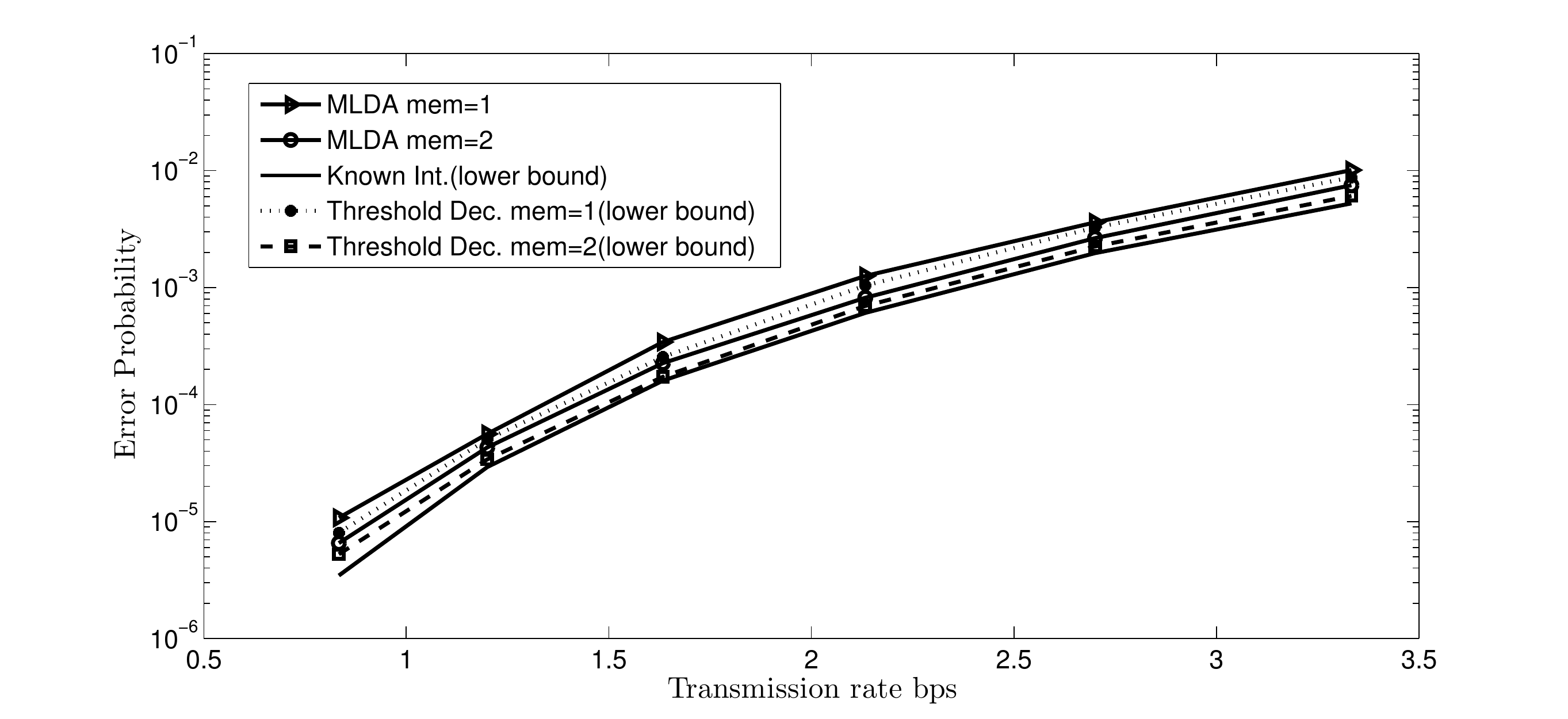}      %%%%%%%%%%%%%%%%%%%% 6 
\label{fig:all_2}
}
%\vspace{-1cm}
\caption{\footnotesize{(a): Main Model (CSK), (b): MCSK. Lower and upper bounds 
for limited memory receivers.}
\iffalse The curves labeled as ``Threshold Dec." are lower bounds on $E_T(R, s, B)$, and ones labeled as ``Known Int." are the lower bounds on $E(R, s, B)$.  Parameters for this figure: $s=60$ and $\lambda_0 =0.4$ 
Not revised\color{black}\fi
}
\label{fig:CSK2}
\vspace{-0.5cm}
\end{figure}

More generally we can compare the lower and upper bounds for the practical
range of $s\in [20:100]$ and $R\in [1:15]bps$ and the background noise $\lambda_0=0$.
Here are the results for $\msP_\msT(R, s, B)$:
\begin{itemize}
\item For $B=1$, the upper bound (\color{black}$\sMLDA(R,s,B)$\color{black}) is less than 1.6 times the lower bound (\color{black}$\sGA_\msT(R,s,B)$\color{black}) throughout
the range of parameters. Therefore we can find $\msP_\msT(R,s, 1)$ that lies
between the two within a factor of $1.6$. So for instance if the error probability
of MLDA is $1.6\times 10^{-3}$, the lower bound (\color{black}$\sGA_\msT(R,s,B)$\color{black}) is greater than
$10^{-3}$ and hence $\msP_\msT(R,s, 1)=a\times 10^{-3}$ for some $a\in [1:1.6]$.
\item For $B=2$, the upper bound (\color{black}$\sMLDA(R,s,B)$\color{black}) is less than 4.3 times the lower bound (\color{black}$\sGA_\msT(R,s,B)$\color{black}) throughout
the range of parameters
\end{itemize}
And the results for $\msP (R, s, B)$ is as follows: for $B=1$, the upper bound (\color{black}$\sMLDA(R,s,B)$\color{black}) is less than 12.5 times the lower bound(\color{black}$\sGA(R,s)$\color{black}). For $B=2$, the upper bound (\color{black}$\sMLDA(R,s,B)$\color{black}) is less than 10.5 times the lower bound (\color{black}$\sGA(R,s)$\color{black}. Observe that a ratio less than or equal to 10 implies that the lower and upper bound are of the same order.

As another example, if we vary the noise $\lambda_0$ in the interval $[0,50]$, $s\in[20,80]$ and $R\in [2,15]bps$, we get the following ratios for $\msP_T(R, s, B)$: For $B=1$, the upper bound (\color{black}$\sMLDA(R,s,B)$\color{black}) is less than 1.75 times the lower bound(\color{black}$\sGA_\msT(R,s,B)$\color{black}). For $B=2$, the upper bound (\color{black}$\sMLDA(R,s,B)$\color{black}) is less than 2.5  times the lower bound(\color{black}$\sGA_\msT(R,s,B)$\color{black}). The results for $\msP (R, s, B)$ is as follows: For $B=1$, the upper bound  (\color{black}$\sMLDA(R,s,B)$\color{black}) is less than 21 times the lower bound(\color{black}$\sGA(R,s)$\color{black}.
For $B=2$, the upper bound (\color{black}$\sMLDA(R,s,B)$\color{black}) is less than 5.5 times the lower bound(\color{black}$\sGA(R,s)$\color{black}. These ratios become much better if we apply the lower and upper bounds to
the MCSK modulation scheme (see Section \ref{sec:MCSK}).

%3-d fig explanation\color{black}

\vspace{-0.4cm}
\section{Proposed Receiver: multiple reads}\label{sec:CB}

We now turn to the second main building block of a general decoder as depicted
in Fig. \ref{fig:transmitterschematic}, namely the counter block. Unlike previous
works, we assume that the value of the counter is read every $t_s/N$
seconds rather than every $t_s$ seconds where $t_s$ is the duration of each
transmission time slot; we call  this a multi-read system.\footnote{
Uniform sampling is not necessarily the best strategy, but it is a simple and conventional
one.
} Here $N$ is a natural number, indicating the number of sub-time-slots.
The number of received molecules in sub-time-slot $q$ of time slot $i$ is
denoted by $Y_{i,q}, q\in [1:N]$.

Similar to \eqref{eqn:AA1}, we can derive probability
of hitting of a molecule in the sub-time-slots. 
Consider a particular molecule released at the beginning of a time slot.
The probability that it hits the receiver during the $q$-th sub-time-slot
of the next $k$-th time slot is equal to
\begin{align}
p_{k,q}=2Q\bigg(\frac{\rho}{\sqrt{(k-1+\frac qN)t_s}}\bigg)-2Q\bigg(\frac{\rho}{\sqrt{(k-1+\frac
{q-1}{N})t_s}}\bigg), \quad k\in\mathbb{N}, q\in [1:N].\label{eqn:AA2}
\end{align}
Note that $\sum_{q}p_{k,q}=p_k$ as defined by \eqref{eqn:AA1}. The key
observation is the fact that for a fixed value of $k$, the hitting probability
$p_{kq}$ varies for different values of $q$, i.e. in general $p_{k,q}\neq
p_{k,q'}$ for $q\neq q'$.  {Therefore, we will have N independent Poisson distributions with \emph{distinct means}. The sum of these $N$ Poisson variables would be the total read in the entire time slot, and would not be a sufficient statistic for decoding the message since the parameters of the $N$ Poisson distributions are not equal. Thus, more samples results in more information and better performance.}

The MLDA decoder for multiple reads is provided next.
As before, we assume that the decoder has a memory of $B$ bits that stores
the last $B$ decoded bits. Using this memory the decoder estimates $\hat{I}_{i,q}$
of time slot $i$ resulted by previous symbols in $q$-th sub-time-slot as
\begin{align}
\hat{I}_{i,q}=\lambda_0 t_s + \sum_{j=1}^{B}{p_{{j+1},q}\hat{X}_{i-j}} +
\frac{(s_0+s_1)}{2}  \sum_{j=B+1}^{\infty}{p_{{j+1},q}}. \label{eqn:Analys1}
\end{align}

The decoder assumes the previously decoded symbols are correctly decoded
and uses the $N$  observations in the current time slot for decoding.
The associated decision rule is as follows:
\begin{align}
 \max_{j\in\{0,1\}}{p(y_{i,1},...,y_{i,N}|X_i=s_j,\hat{X}_{i-1}=\hat{x}_{i-1},...,\hat{X}_{i-B}=\hat{x}_{i-B})},
\end{align}
which results in
\begin{align}\label{DR}
\sum_{q=1}^{N}{y_{i,q} \ln {(\frac{\hat{I}_{i,q}+p_{1,q}s_1}{\hat{I}_{i,q}+p_{1,q}s_0})}}\gtrless
^{s_1}_{s_0} (s_1-s_0)\sum_{q=1}^{N}{p_{1,q}} = (s_1-s_0)p_1.
\end{align}

We observe that if the probabilities of hitting for all sub-time
slot were the same, the above multiple reads decision rule reduces to the
one read decision rule. The analysis of the probability of error of the above decoder using the saddle-point method is given in Appendix \ref{appndxC}.

To justify receiver sampling at a higher rate than the transmission
rate, we numerically examine the performance of multiple-read MLDA.  We see that the performance improves by increasing
$N$, significantly for small values of $N$, but the amount of increase saturates from some
point onwards. 

%Fig. \ref{fig9} demonstrates the effect of multi-sampling on the probability of error in MCSK Modulation for $s=800$, $\lambda_0=40$, and $B=3$. It can be concluded that by increasing the number of sampling in each time slot, probability of error decays.

Fig. \ref{figNR} shows the probability of error of MLDA versus time slot durations for different number of observations in one time slot resulted by simulation and also saddle point analysis. The figure shows the analysis is compatible with the simulation results. Also, it is observed that increasing $N$, significantly improves the probability of error. Specially, as  $t_s$ increases the improvement is more noticeable.
Fig. \ref{figBN} shows the probability of error of MLDA versus $N$ for different values of decoder memory, $B$, computed by the saddle point analysis. It is observed that increasing $N$ up to 4  significantly improves the probability of error. For $N>4$ the improvement is not noticeable.

%Fig. \ref{figNS} depicts the probability of error of MLDA versus the parameter $s$ (transmission power is $s/2$) for different values of $N$. It is observed, the improvement resulted by more number of observations is more as $s$ increases.

\vspace{-0.4cm}
\section{Application to MCSK}\label{sec:MCSK}

MCSK is a modulation technique introduced in \cite{AA1} in which two molecule types are utilized, {though not for data transmission.} \emph{At the transmitter}, the molecule types (for instance, $A_1$ and $A_2$) alternate from one symbol interval to the next one (so for instance molecule type $A_1$ is used in odd time slots and molecule type $A_2$ is used in even time slots). This causes a reduction in the interference from the same molecule type. More specifically, if we focus on molecules of type $A_1$, we would have $X_{i}=0$ for all even $i$ (no transmission of $A_1$ molecules in the even time slots). Therefore the interference plus noise term would become 
$I_i=\lambda_0 t_s + \sum_{k=1}^\infty p_{k+1} X_{i-k}= \lambda_0 t_s + \sum_{k=2, k~even}^\infty p_{k+1} X_{i-k}$. The ideas developed above can be adapted to MCSK. For numerical evaluations, we have assumed that $\rho^2=0.3$. In all of the figures we assumed $B$ bits for storing the data of each molecule type in the MLDAs (thus $2B$ bits of memory in total).  Fig. \ref{fig4} depicts the plot of probability of error for MCSK scheme for proposed simple receiver for different values of $B$ versus transmission rate for $s=60$ and background noise power $\lambda_0=0.4$. In the same figure, probability of error of an ML decoder with  \emph{known interference} (interference is fully available at the receiver) is drawn. This curve clearly serves as a lower bound. It is seen that as the memory $B$ increases, probability of error of the MLDA decays. Further, the figure shows that for $B=3$ the simple receiver is pretty close to the performance of the  ML decoder with known interference. This figure also contains a curve labeled by" the int. is not considered", which corresponds to a ``conventional receiver" that ignores the existence of interference, but uses MAP decoding. This simple decoder has been commonly used in previous works; for instance see \cite{AA2}.\color{black}

Fig. \ref{fig5} depicts the probability of error for proposed simple receivers versus the transmission rate for $s=80$ and  background noise power $\lambda_0=40$ for MCSK modulation. It is observed that as the transmission rate increases ($t_s$ decreases), the environmental noise decreases as well, while the main signal term and the interference term increase. Accordingly, there is a trade-off between the constructive and destructive terms. It can be interpreted from the figure that by increasing the transmission rate, the probability of error first decreases until it reaches its minimum value for a specific value of transmission rate (about 1.5 depending on the curve). From here increasing the rate increases all of probabilities of error.

\iffalse
Fig. \ref{fig7} is same as Fig. \ref{fig:CSK2} except for MCSK modulation. It can be concluded that the derived simple receivers for this modulation are very close to the lower bounds.
\fi

 Fig. \ref{fig7_mid} plots the maximum possible rate when the probability of error is less than or equal to 0.005 for a given $s$. This is plotted for the known interference lower bound and MLDAs with different memories.
  {Fig. \ref{Fig3d} is same as Fig. \ref{fig:CSK2} (a) except that it is for a three dimensional medium; here $s=120$. The transmitter is located at $[0,0,0]^T$; the receiver is assumed to be a cube of length of $2 \mu m$, with the center of the cube located at $[8,8,8]^T \mu m$. Hitting probabilities were estimated using an extensive numerical simulation.}

We end this section by diverging from the main theme of the paper and briefly comment on the MCSK modulation scheme. 
In the original MCSK, it was assumed that the receiver counts the number of molecules of type $A_1$ in odd time slots and molecules of type $A_2$ in even time slots. If it is possible to count both molecule types in the same time slot, one can postpone decoding of 
 the current symbol for one time slot to exploit the molecules counted in both of the current and the next time slot, i.e. $Y_i$ and $Y_{i+1}$ in the decoding process. In other words, in the even time slots where the transmitter releases molecules of type $A_2$, we  count molecules of type $A_1$ that arrive from transmission in the previous time slot.  For the MAP decision rule for one-delay decoding, one has to compare $p(Y_i=y_1, Y_{i+1}=y_1, |X_i=x, K_i=k)$ for different values of $x$. This probability is equal to
$$\mathbb{E}_{I_i, I_{i+1}|K_i=k}\frac{e^{-(p_1x+I_i)}(p_1x+I_i)^{y_1}}{y_1!}\frac{e^{-(p_2x+I_{i+1})}(p_2x+I_{i+1})^{y_2}}{y_2!}$$
where we used the fact that given $I_{i}, I_{i+1}$ and $X$, rv's $Y_i$ and $Y_{i+1}$ are independent because of the thinning property of Poisson distribution. The formula for MLDA decoder with one delay for the MCSK can be computed as follows: having 
$K_i=(\hat{X}_{i-2}, \hat{X}_{i-4}, \cdots,
\hat{X}_{i-2B})$  in the memory, we estimate 
\begin{align*}
\hat{I}_i&=\lambda_0t_s+p_3\hat{X}_{i-2}+\cdots+p_{2B+1}\hat{X}_{i-2B}+\frac {s_0+s_1}{2}\sum_{j=B+1}^{\infty}
p_{2j+1},\\
\hat{I}_{i+1}&=\lambda_0t_s+p_4\hat{X}_{i-2}+\cdots+p_{2B+2}\hat{X}_{i-2B}+\frac {s_0+s_1}{2}\sum_{j=B+1}^{\infty}
p_{2j+2}.
\end{align*} Then the decision rule can be worked out as follows: 
$$\argmin_{x\in\{s_0, s_1\}}e^{-(p_1+p_2)x}(p_1x+\hat{I}_i)^{y_1}(p_2x+\hat{I}_{i+1})^{y_2}=\argmin_{x\in\{s_0, s_1\}}-(p_1+p_2)x+y_1\log(p_1x+\hat{I}_i)+y_2\log(p_2x+\hat{I}_{i+1}),$$
resulting in a threshold rule. For MLDA with one delay, when having multi-read, a decision rule similar to \eqref{DR} can be obtained.
 Fig. \ref{fig12} demonstrates both effects of delay decoding and multi-sampling. It is seen that with large transmission rates, probability of error decreases when we use delay decoding plus multi-sampling.

\iffalse
Fig. \ref{fig10} demonstrates the effect of delay decoding, showing that waiting for one time slot can improve the performance of receivers. Fig. \ref{fig11} is same as Fig \ref{fig4} but for a constant value of $t_s$ versus different values of average transmission power, $\frac{s}{2}$. It depicts the probability of error of the zero delay and delayed decoding in two cases of with and without considering interference in terms of the average transmission power for the symbol duration of $t_s=150ms$. It is observed that not considering the interference effect will devastate the receiver performance. It is also recognizable that using $Y_{i+1}$ in delayed decoding has improved the performance; while in the case of zero-delay decoding that does not use $Y_{i+1}$, the interference is not well managed.

Fig. \ref{fig12} demonstrates both effects of delay decoding and multi-sampling. It is seen that in large transmission rates, probability of error decreases when we use delay decoding plus multi-sampling.

\fi
\vspace{-0.7cm}
\section{Conclusions}\label{sec:conclusion}
In this paper, we studied diffusion-based molecular communication and demonstrated that we can use a simple threshold decoder to attain error probabilities close to optimal ones. This is of practical importance since nanomachines are resource-limited devices. We further proved a theoretical result that reduces the computational complexity of finding the performance of an optimal threshold decoder, and found a sufficient condition under which an optimal ML decoder becomes a threshold one. \color{black} We also studied the effect of multiple reads at the receiver, and showed that taking multiple samples can improve the performance significantly. As a future work, one can relax the assumptions we have made (uncoded transmission at the transmitter and the zero delay at the receiver)  {and propose  simple, practical transmitter and receivers that are near optimal for more general class of encoder and decoders}. We believe that the ideas introduced in this paper, including the genie-aided lower bounds, can be helpful in that quest.

\appendix
\vspace{-0.34cm}
\subsection{Proof of Theorem \ref{thm:thm3}}

\begin{proof}Recall that we assume a genie  provides $B$ bits to the decoder. The genie maps  $I_i = \alpha$ to the value $k$ where $k \in\{1,2,\cdots, 2^B\}$. With a slight abuse of notation, we also denote this mapping by $k(\alpha)$. The decoder in turn maps $k$ to the employed threshold, $\tau_k\in\mathbb{Z}$; thus the interference $ \alpha$ would end up mapping to $\tau_{k(\alpha)}$.

For a given value of interference $\alpha$ and an arbitrary threshold $T$, the probability of error for equally likely transmission rates $X_i$ is
 \begin{align}
\mP(E | \alpha,T) =\frac{1}{2} \left[\sum_{y< T} e^{-(\alpha+p_1s)} \frac{(\alpha+p_1s)^y}{y!}
+ \sum_{y \geq T} e^{-\alpha} \frac{\alpha^y}{y!} \right]\label{eq:ap2}
\end{align}
Thus total probability of error, averaging over the interference is,
\begin{align}\mP(E) = \sum_{\alpha}\mathbb{P}\left[I_i=\alpha \right]\mP(E|\alpha, \tau_{k(\alpha)}).\label{eq:ap1}
\end{align}
To minimize $\mP(E)$ one needs to do minimization over two sets (1) minimization over the genie's mapping, $k(\alpha)$, and (2) minimization over the set of thresholds $\{\tau_k, 1\leq k\leq 2^B\}$. We can first do minimization over (2) and then over (1), i.e.
$\min_{\{\tau_k, 1\leq k\leq 2^B\}}\min_{k(\alpha)}\mP(E).$
We prove a much stronger statement than  needed. We show that for \emph{any} fixed choice of  $\{\tau_k, 1\leq k\leq 2^B\}$, when we minimize probability of error over set (1) of all mappings $k(\alpha)$, the form of the solution is one of quantizing the interference values into $2^B$ bins and using the interference interval's bin index as the value for $k(\alpha)$. Take a fixed set of thresholds $\{\tau_k, 1\leq k\leq 2^B\}$. Using \eqref{eq:ap1}, to minimize the
probability of error, the genie should choose $k(\alpha)$ as follows:
\begin{align}k(\alpha)=\arg\min_{k\in [1:2^B]}\mP(E|\alpha,\tau_{k}).\label{eqn:alphaMAP}\end{align}
The optimizing $k$ may not be unique, in which case, we can choose any value from the optimizing set.
We wish to show that the mapping $\alpha\mapsto k(\alpha)$ results in the quantization of the interference as stated in the theorem.  
It suffices for us to show that if $\alpha_1$ and $\alpha_2$ are mapped to
some $k^*$, and $\alpha_3$ is in the interval $[\alpha_1, \alpha_2]$ then $\alpha_3$
can be also mapped to the same $k^*$. In other words, for all $k'$ we have
$\mP(E|\alpha_3,\tau_{k^*})\leq \mP(E|\alpha_3,\tau_{k'}).$ We show our desired statement via contradiction.
Suppose there is some $k'\neq k^*$ such that
$\mP(E|\alpha_3,\tau_{k^*})> \mP(E|\alpha_3,\tau_{k'}).$ Since $\alpha_1$ and $\alpha_2$ were mapped to $k$, by equation \eqref{eqn:alphaMAP},
we get that
$\mP(E|\alpha_1,\tau_{k^*})\leq \mP(E|\alpha_1,\tau_{k'})$ and $\mP(E|\alpha_2,\tau_{k^*})\leq \mP(E|\alpha_2,\tau_{k'}).$
Consider the difference function $d(x)=\mP(E|x,\tau_{k'})- \mP(E|x,\tau_{k^*})$, where by $\mP(E|x,\tau_{k'})$ we mean the right hand side of \eqref{eq:ap2} when we formally replace $\alpha$ with a variable $x$, and $T$ with $\tau_{k'}$. Then $d(\alpha_1)\geq
0$, $d(\alpha_2)\geq 0$ and $d(\alpha_3)<0$. Since $0\leq \alpha_1<\alpha_3<\alpha_2$,
and $d(x)$ is continuous in $x$, the function $d(x)$
would have at least two non-negative zeros. 

We will arrive at a contradiction by showing that $d(x)$ has at most one
non-negative zero. Since the number of non-negative zeros of $d(x)$ and $-d(x)$
are the same, without loss of generality we assume that $\tau_{k'}> \tau_{k^*}$. Let
$T_1=\tau_{k^*}, T_2=\tau_{k'}$. Then
$d(x)=\frac 12 \sum_{y=T_1}^{T_2-1}\frac{e^{-x}}{y!}\left [ x^y-e^{-p_1s}(x+p_1s)^y\right].$
The number of non-negative zeros of $d(x)$ is the same as the number of non-negative
zeros of $e^xd(x)$. Observe that 
$e^xd(x)=\frac 12 \sum_{y=T_1}^{T_2-1}\frac{1}{y!}\left [ x^y-e^{-p_1s}(x+p_1s)^y\right]$
is a polynomial. Since zero is not a root of this polynomial, non-negative
roots are equivalent to positive roots. Thus  we can use Descartes' rule
of signs to find an upper bound on the number of positive zeros. According
to Descartes' rule of signs, the number of sign changes between consecutive
non-zero coefficients of a polynomial is an upper bound on the number of positive
zeros. Expanding the terms in $e^xd(x)=\sum_{n=0}^{T_2-1}a_n x^n$ where
\begin{align}
a_n&=-\frac 12 e^{-p_1s}\sum_{y=T_1}^{T_2-1}\frac{1}{y!}{y \choose n}(p_1s)^{y-n},
& n<T_1
\\
a_n&=\frac 12 \left[ \frac{1}{n!}- e^{-p_1s}\sum_{y=n}^{T_2-1}\frac{1}{y!}{y
\choose n}(p_1s)^{y-n} \right], & n\in [T_1:T_2-1]
\end{align}
Clearly $a_n< 0$ for $n<T_1$. We claim that $a_n>0$ for $n\in [T_1:T_2-1]$,
implying that there is no more than one change of sign. For $ n\in [T_1:T_2-1]$
we have
\begin{align*}
a_n&=\frac 12 \left[\frac{1}{n!}- e^{-p_1s}\sum_{y=n}^{T_2-1}\frac{1}{y!}{y
\choose n}(p_1s)^{y-n}\right]
=\frac{1}{2(n!)} \left[1- e^{-p_1s}\sum_{y=0}^{T_2-n-1}\frac{1}{y!}(p_1s)^{y}\right]
\end{align*}
which is positive since $e^{p_1s}=\sum_{y=0}^{\infty}\frac{1}{y!}(p_1s)^{y}>\sum_{y=0}^{T_2-n-1}\frac{1}{y!}(p_1s)^{y}.$
\end{proof}

\vspace{-0.7cm}
\subsection{Proof of Theorem \ref{thm:thm2}}
\begin{proof}
Consider a zero-delay memory limited decoder as defined in subsection \ref{definition:mlzd},
with the parameters $s, B$ and $R$. Independent symbols
$X_1, X_2, \cdots$ are transmitted. Further, assume that the decoder uses
the optimal MAP decoding rule for determining $\hat{X}_i$, {\em  i.e.}
with $K_i=k$ stored in memory, at time $i$, the decoder compares $\mathbb{P}(Y_i=y|K_i=k,X_i=0)$
and $\mathbb{P}(Y_i=y|K_i=k,X_i=s)$.\footnote{Note that $K_i$ is independent of $X_i$;
 we have also assumed equal probability messages, and therefore the MAP and ML decision
rules are the same.}.  Let us define
\begin{align}
P_{0k}(y) &= \mathbb{P}(Y_i=y|K_i=k,X_i=0) = \sum_{\alpha}\mathbb{P}(I_i=\alpha|K_i=k)
e^{-\alpha} \frac{\alpha^y}{y!}, \nonumber \\ 
P_{1k}(y) &= \mathbb{P}(Y_i=y|K_i=k,X_i=s) = \sum_{\alpha}\mathbb{P}(I_i=\alpha|K_i=k)e^{-(\alpha+p_1s)}\frac{(\alpha+p_1s)^y}{y!},
\end{align}
where $I_i=\lambda_0 t_s + \sum_{k=1}^\infty p_{k+1} X_{i-k}$ is the interference plus noise
at time $i$ (and is independent of $X_i$). Although not needed in the proof, we note that the probability of error of the
decoder at time $i$ can be written as
$
\mathbb{P}(\hat{X}_i\neq X_i) = \frac{1}{2} \left[\sum_{k,y}\mathbb{P}(K_i=k) \min(P_{0k}(y),P_{1k}(y)) \right].
$
We wish to determine the conditions on the rate $R$ such that optimal MAP decoding reduces to a threshold detector in the presence of intersymbol interference.  Equivalently, what are the conditions on $R$ such that for each $k$, there exists a $\tau(k)$ such that for all $ y<\tau(k)$ we have $\max(P_{0k}(y),P_{1k}(y))=P_{0k}(y)$ and for all $y\geq \tau(k)$ we have $\max(P_{0k}(y),P_{1k}(y))=P_{1k}(y).$
These expressions are in turn, equal to:  
\begin{align}P_{0k}(y)\geq P_{1k}(y)\quad \forall y< \tau(k),\label{eqn:P1}
\qquad \qquad P_{0k}(y)\leq P_{1k}(y)\quad\forall y\geq \tau(k).\end{align}
Let $\alpha_{max}$ and $\alpha_{min}$ be the maximum and minimum possible interference values, respectively, i.e. $\alpha_{min}=\lambda_0 t_s$ and $\alpha_{max}=\lambda_0 t_s + s\sum_{k=1}^\infty p_{k+1}.$ The following constraints form a set of sufficient conditions for equation \eqref{eqn:P1}
 to hold:
(1) $p_1s +\alpha_{min}\geq \alpha_{max}$,~~~ (2) $P_{0k}(y)\geq P_{1k}(y),  \forall y\in[0,\alpha_{max}]$, (3) $P_{0k}(y)\leq P_{1k}(y),  \forall y \geq p_1s+\alpha_{min}$,~~ and (4) $P_{0k}(y)$ is a decreasing function of $y$ and $P_{1k}(y)$ is an
increasing function of $y$ in $[\alpha_{max}, p_1s +\alpha_{min}]$.

The first condition is equivalent to  $p_1\geq \sum_{j=1}^{\infty}p_j/2$.
For the second condition, observe that for any $y\in[0,\alpha_{max}]$,
\begin{align*}P_{1k}(y)&= \sum_{\alpha}\mathbb{P}(I_i=\alpha|K_i=k)e^{-(\alpha+p_1s)}\frac{(\alpha+p_1s)^y}{y!}
= \sum_{\alpha}\mathbb{P}(I_i=\alpha|K_i=k)e^{-\alpha}\frac{\alpha^y}{y!}e^{-p_1s}\left(\frac{\alpha+p_1s}{\alpha}\right)^y
\\&\leq \sum_{\alpha}\mathbb{P}(I_i=\alpha|K_i=k)e^{-\alpha}\frac{\alpha^y}{y!}e^{-p_1s}\left(\frac{\alpha_{min}+p_1s}{\alpha_{min}}\right)^{\alpha_{max}}
=e^{-p_1s}\left(\frac{\alpha_{min}+p_1s}{\alpha_{min}}\right)^{\alpha_{max}}P_{0k}(y)
\end{align*}
Thus the second condition holds if 
$
e^{-p_1s}\left(\frac{\alpha_{min}+p_1s}{\alpha_{min}}\right)^{\alpha_{max}}\leq
1.
$
After simplification we get:
\begin{align}\label{eqn:cons2}
p_1 \geq \frac{\alpha_{max}}{s} \ln\left(\frac{\alpha_{min}+p_1s}{\alpha_{min}}\right)=\frac{\lambda_0
t_s + s\sum_{j=2}^{\infty}p_j}{s} \ln\left(\frac{\lambda_0 t_s + p_1s}{\lambda_0 t_s}\right).
\end{align}
For the third condition, observe that for any $y \geq p_1s+\alpha_{min}$ we have,
\begin{align*}P&_{1k}(y)=\sum_{\alpha}\mathbb{P}(I_i=\alpha|K_i=k)e^{-\alpha}\frac{\alpha^y}{y!}e^{-p_1s}\left(\frac{\alpha+p_1s}{\alpha}\right)^y
\\&\geq \sum_{\alpha}\mathbb{P}(I_i=\alpha|K_i=k)e^{-\alpha}\frac{\alpha^y}{y!}e^{-p_1s}\left(\frac{\alpha_{max}+p_1s}{\alpha_{max}}\right)^{p_1s+\alpha_{min}}
=e^{-p_1s}\left(\frac{\alpha_{max}+p_1s}{\alpha_{max}}\right)^{p_1s+\alpha_{min}}P_{0k}(y).
\end{align*}
Thus the second condition holds if 
$
e^{-p_1s}\left(\frac{\alpha_{max}+p_1s}{\alpha_{max}}\right)^{p_1s+\alpha_{min}}\geq
1.
$
After simplification, we get:
\begin{align}\label{eqn:cons3}
p_1 \leq \frac{p_1s+\alpha_{min}}{s} \ln\left(\frac{\alpha_{max}+p_1s}{\alpha_{max}}\right)=\frac{\lambda_0
t_s + p_1s}{s} \ln\left(\frac{\lambda_0 t_s + s\sum_{j=1}^{\infty}p_j}{\lambda_0 t_s + s\sum_{j=2}^{\infty}p_j
}\right).
\end{align}
The fourth condition automatically holds, since a Poisson
distribution with parameter $\lambda$ is increasing for $0\leq y\leq \lfloor\lambda\rfloor$
and decreasing for $y\geq \lfloor\lambda\rfloor$. Since $P_{0k}(y)$ and $P_{1k}(y)$
are linear combinations of several Poisson distributions with means $\alpha$
and $\alpha+p_1s$ respectively, $P_{0k}(y)$ is decreasing after the maximum
value of $\alpha$, i.e. $\alpha_{max}$ and $P_{1k}(y)$ is increasing before
the minimum value of $\alpha+p_1s$, i.e. $\alpha_{min}+p_1s$.

Therefore, we need to find constraints on $s$ that would imply  $p_1\geq \sum_{j=1}^{\infty}p_j/2$ and equation
\eqref{eqn:cons3}. Substituting $t_s=1/R$ we can rewrite these
equations  as follows:
\begin{align*}
\frac{\lambda_0 + Rp_1s}{Rs} \ln\left(\frac{\lambda_0+ Rs\sum_{j=1}^{\infty}p_j}{\lambda_0+ Rs\sum_{j=2}^{\infty}p_j}\right)\geq
p_1 \geq \max\left[\frac{\sum_{j=1}^{\infty}p_j}{2}, \frac{\lambda_0 + Rs\sum_{j=2}^{\infty}p_j}{Rs} \ln\left(\frac{\lambda_0
+ Rp_1s}{\lambda_0}\right)\right].
\end{align*}
If we let $s$ converge to infinity, we get $p_1\ln\left(\sum_{j=1}^{\infty}p_j/\sum_{j=2}^{\infty}p_j\right)>p_1>\sum_{j=1}^{\infty}p_j/2$ that is satisfied as long as $p_1>(e-1)\sum_{j=2}^{\infty}p_j$.
\end{proof}

\iffalse
\emph{Derivation of the Simple Decoder for Multiple Reads: } 

As before we assume that the decoder has a memory of $B$ bits that stores the last $B$ decoded bits. Using this memory the decoder estimates the interference of time slot $i$ resulted by previous symbols in $q$-th sub-timeslot as

\begin{align}
\hat{I}_{i,q}=\lambda_0 t_s + \sum_{j=1}^{B}{p_{{j+1},q}\hat{X}_{i-j}} + \frac{(s_0+s_1)}{2}  \sum_{j=B+1}^{\infty}{p_{{j+1},q}}. \label{eqn:Analys1}
\end{align}

 The decoder assumes the previously decoded symbols are correctly decoded and uses the $N$ taken observations in the current time slot for decoding. The decision rule of this suboptimal MAP decoder is obtained as follows:
\begin{align}
 \max_{j\in\{0,1\}}{p(y_{i,1},...,y_{i,N}|X_i=s_j,\hat{X}_{i-1}=\hat{x}_{i-1},...,\hat{X}_{i-B}=\hat{x}_{i-B})},
\end{align}
which results in
\begin{align}\label{DR}
\sum_{q=1}^{N}{y_{i,q} \ln {(\frac{\hat{I}_{i,q}+p_{1,q}s_1}{\hat{I}_{i,q}+p_{1,q}s_0})}}\gtrless ^{s_1}_{s_0} (s_1-s_0)\sum_{q=1}^{N}{p_{1,q}} = (s_1-s_0)p_1.
\end{align}

 It is worth to note that if the probabilities of hitting for all sub-time slot were the same, the above multiple reads decision rule reduces to the one read decision rule.
\fi

\vspace{-0.5cm}
\subsection{Analysis of Probability of Error of Mutli-Read MLDA}\label{appndxC}
As the decoding of the current symbol is dependent on both the last transmitted and decoded symbols, $\mathbb{P}(E|s_0)$ and $\mathbb{P}(E|s_1)$ can be computed recursively from a system of equations similar to that explained earlier. To obtain this system of equations, one needs to compute the probabilities $$\mathbb{P}(E|X_i=x_i,\hat{I}_{i,1}=\hat{\alpha}_1, \cdots ,\hat{I}_{i,N}=\hat{\alpha}_N, I_{i,1}=\alpha_1,\cdots , I_{i,N}=\alpha_N).$$ Let us define the rv $Z=\sum_{q=1}^{N}{Y_{i,q} \ln {(\frac{\hat{I}_{i,q}+p_{1,q}s_1}{\hat{I}_{i,q}+p_{1,q}s_0})}}=\sum_{q=1}^{N}{Y_{i,q}A_q}$ and $\tau=(s_1-s_0)p_1$. Then we have
\begin{align}
&\mathbb{P}(E|X_i=s_1,\hat{I}_{i,1}=\hat{\alpha}_1, \cdots ,\hat{I}_{i,N}=\hat{\alpha}_N, I_{i,1}=\alpha_1,\cdots , I_{i,N}=\alpha_N)\nonumber\\
&=\mathbb{P}(Z<\tau|X_i=s_1,\hat{I}_{i,1}=\hat{\alpha}_1, \cdots ,\hat{I}_{i,N}=\hat{\alpha}_N, I_{i,1}=\alpha_1,\cdots , I_{i,N}=\alpha_N)\nonumber\\&\qquad+\frac 12 \mathbb{P}(Z=\tau|X_i=s_1,\hat{I}_{i,1}=\hat{\alpha}_1, \cdots ,\hat{I}_{i,N}=\hat{\alpha}_N, I_{i,1}=\alpha_1,\cdots , I_{i,N}=\alpha_N)
\end{align}
and
\begin{align}
&\mathbb{P}(E|X_i=s_0,\hat{I}_{i,1}=\hat{\alpha}_1, \cdots ,\hat{I}_{i,N}=\hat{\alpha}_N, I_{i,1}=\alpha_1,\cdots , I_{i,N}=\alpha_N)\nonumber\\
&=\mathbb{P}(Z>\tau|X_i=s_0,\hat{I}_{i,1}=\hat{\alpha}_1, \cdots ,\hat{I}_{i,N}=\hat{\alpha}_N, I_{i,1}=\alpha_1,\cdots , I_{i,N}=\alpha_N)\nonumber\\&\qquad+\frac 12 \mathbb{P}(Z=\tau|X_i=s_0,\hat{I}_{i,1}=\hat{\alpha}_1, \cdots ,\hat{I}_{i,N}=\hat{\alpha}_N, I_{i,1}=\alpha_1,\cdots , I_{i,N}=\alpha_N).
\end{align}

Since the distribution of the decision rule in (\ref{DR}), which is a weighted sum of Poisson variables, does not have a closed form, we use Saddle Point approximation  \cite{saddle} to analytically evaluate the performance. Let us denote the characteristic function of a rv $Z$ as $\Phi_{Z}(\zeta)=\mathbb{E}(e^{\zeta Z})$. Based on the saddle point approximation we have
\begin{align}\label{mod1}
&\mathbb{P}(Z>\tau|x_i,\hat{\alpha}_{1},...,\hat{\alpha}_{N},\alpha_{1},...,\alpha_{N})+\frac 12 \mathbb{P}(Z=\tau|x_i,\hat{\alpha}_{1},...,\hat{\alpha}_{N},\alpha_{1},...,\alpha_{N})=
\nonumber\\&\qquad \frac{\exp(\Psi_{Z|x_i,\hat{\alpha}_{1},...,\hat{\alpha}_{N},\alpha_{1},...,\alpha_{N}}(\zeta_0))}{\sqrt{2\pi\Psi^{''}_{Z|x_i,\hat{\alpha}_{1},...,\hat{\alpha}_{N},\alpha_{1},...,\alpha_{N}}(\zeta_0)}}(1+R(\zeta_0)),
\end{align} 
\begin{align}\label{mod2}
&\mathbb{P}(Z<\tau|x_i,\hat{\alpha}_{1},...,\hat{\alpha}_{N},\alpha_{1},...,\alpha_{N})+\frac 12 \mathbb{P}(Z=\tau|x_i,\hat{\alpha}_{1},...,\hat{\alpha}_{N},\alpha_{1},...,\alpha_{N})=
\nonumber\\&\qquad \frac{\exp(\Psi_{Z|x_i,\hat{\alpha}_{1},...,\hat{\alpha}_{N},\alpha_{1},...,\alpha_{N}}(\zeta_1))}{\sqrt{2\pi\Psi^{''}_{Z|x_i,\hat{\alpha}_{1},...,\hat{\alpha}_{N},\alpha_{1},...,\alpha_{N}}(\zeta_1)}}(1+R(\zeta_1)),
\end{align}
 in which
$
\Psi_{Z|x_i,\hat{\alpha}_{1},...,\hat{\alpha}_{N},\alpha_{1},...,\alpha_{N}}(\zeta)=\ln\left(\frac{\Phi_{Z|x_i,\hat{\alpha}_{1},...,\hat{\alpha}_{N},\alpha_{1},...,\alpha_{N}}(\zeta)e^{-\tau \zeta}}{|\zeta|}\right)$,
and $\zeta_0$ and $\zeta_1$ are the positive and negative roots of the following equation, respectively:
\begin{equation}\label{mod3}
\frac{d\Psi_{Z|x_i,\hat{\alpha}_{1},...,\hat{\alpha}_{N},\alpha_{1},...,\alpha_{N}}(\zeta)}{d\zeta}=0.
\end{equation}
Observation $Y_{i,q}$ conditioned to $X_i=x_i,\hat{I}_{i,1}=\hat{\alpha}_1,...,\hat{I}_{i,N}=\hat{\alpha}_N,I_{i,1}=\alpha_1,...,I_{i,N}=\alpha_N$ is a Poisson rv:
$
Y_{i,q} \sim Poisson(\lambda_q)
$
where $\lambda_q=\hat{\alpha}_q+p_{1,q}x_i$ and therefore
$
\Phi_{Y_{i,q}|x_i,\hat{\alpha}_{1},...,\hat{\alpha}_{N},\alpha_{1},...,\alpha_{N}}(\zeta)=e^{(\lambda_q(e^\zeta-1))}
$. On the other hand 
\begin{align*}
\Phi_{Z|x_i,\hat{\alpha}_{1},...,\hat{\alpha}_{N},\alpha_{1},...,\alpha_{N}}(\zeta)&=\mathbb{E}\{e^{\zeta Z}|x_i,\hat{\alpha}_{1},...,\hat{\alpha}_{N},\alpha_{1},...,\alpha_{N}\}=\mathbb{E}\{e^{\zeta\sum_{q=1}^{N}{A_qY_{i,q}}}|x_i,\hat{\alpha}_{1},...,\hat{\alpha}_{N},\alpha_{1},...,\alpha_{N}\}
\\
&=\prod_{q=1}^{N}{\mathbb{E}\{e^{\zeta A_qY_{i,q}}|x_i,\hat{\alpha}_{1},...,\hat{\alpha}_{N},\alpha_{1},...,\alpha_{N}\}}
=\prod_{q=1}^{N}{\phi_{Y_{i,q}|x_i,\hat{\alpha}_{1},...,\hat{\alpha}_{N},\alpha_{1},...,\alpha_{N}}(A_q\zeta)}
\\
&=e^{\sum_{q=1}^{N}{\lambda_q(e^{\zeta{A_q}}-1)}}.
\end{align*}
Therefore, we have\small
\begin{equation}
\Psi_{Z|x_i,\hat{\alpha}_{1},...,\hat{\alpha}_{N},\alpha_{1},...,\alpha_{N}}(\zeta)=\log\left(\frac{e^{\sum_{q=1}^{N}{\lambda_q(e^{\zeta{A_q}}-1)-\zeta (s_1-s_0) p_{1,q}}}}{|\zeta|}\right)=\sum_{q=1}^{N}{\lambda_q(e^{\zeta{A_q}}-1)-\zeta (s_1-s_0) p_{1,q}}-\ln(\zeta).
\end{equation}\normalsize
Now, we compute the roots of (\ref{mod3}):
$
\frac{d\Psi_{Z|x_i,\hat{\alpha}_{1},...,\hat{\alpha}_{N},\alpha_{1},...,\alpha_{N}}(\zeta)} {d\zeta}=\sum_{q=1}^{N}{A_q\lambda_qe^{\zeta A_q}-p_{1,q} (s_1-s_0)}-\frac{1}{\zeta}=0.
$
This equation can be solved numerically using, for example, Newton's algorithm. For the case $X_i=s_0$, $\zeta_e$ is expected be negative and therefore in the Newton algorithm the initial value of root should be taken to be negative. In this case (\ref{mod2}) can be used to compute the probability of error conditioned to $X_i=s_0$ by taking average over the possible values of the interference terms. For the case $X_i=s_1$, $\zeta_e$ is expected be positive and hence in the Newton algorithm a positive initial value of root can be chosen. Equation (\ref{mod1}) can be used to compute the probability of error conditioned to $X_i=s_1$ by taking average over the possible values of the interference terms. The term of $1+R(\zeta)$ can be approximated with 1 (\cite{saddle}).

%%%%%%%%%%%%%%%%%%%%%%%%%%%%%%%%%%%%%
\newpage
\begin{figure}
\begin{center}
\includegraphics[width=170mm]{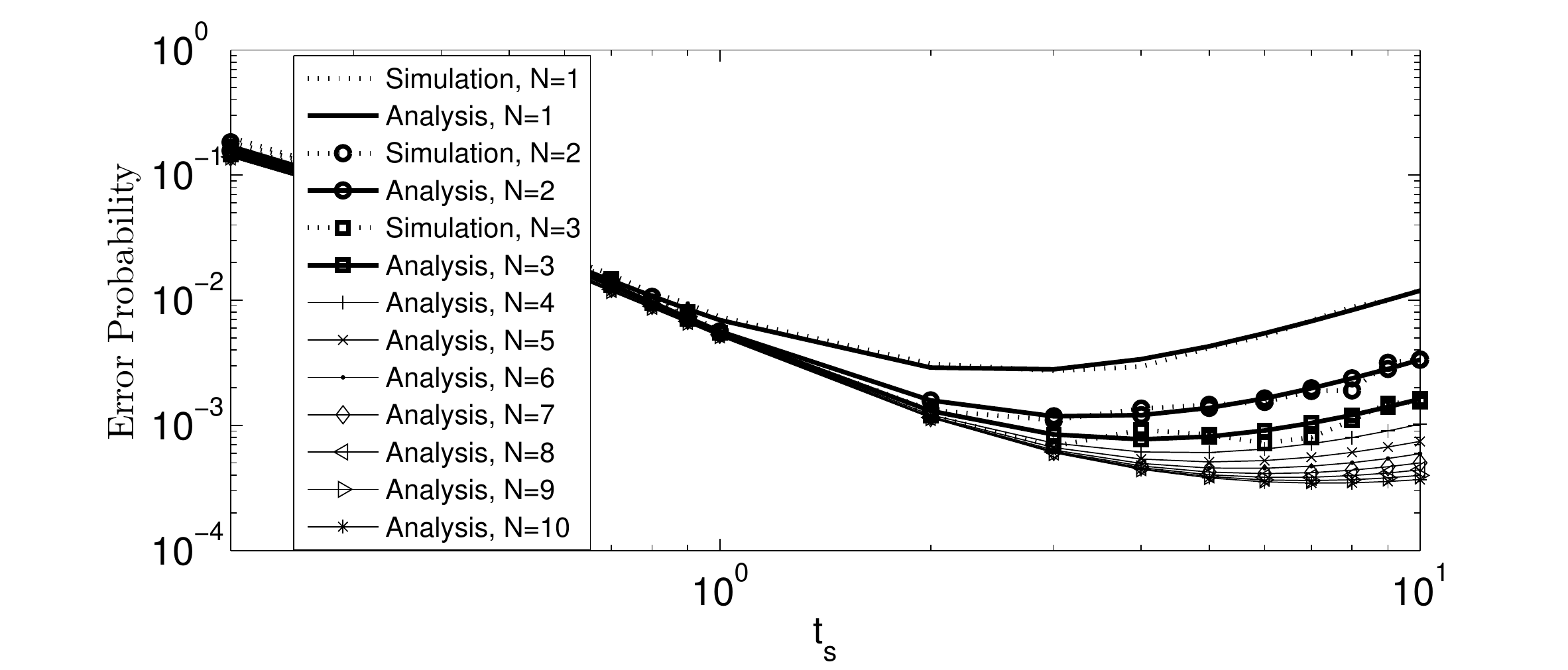}
\end{center}
\vspace{-1cm}
\caption{Probability of error for zero delay decoder with multiple reads in terms of time slot duration. Parameters for this figure: $B=1$, $s=40$ and $\lambda_0 = 4$}
\label{figNR}
\end{figure}

\begin{figure}
\begin{center}
\includegraphics[width=170mm]{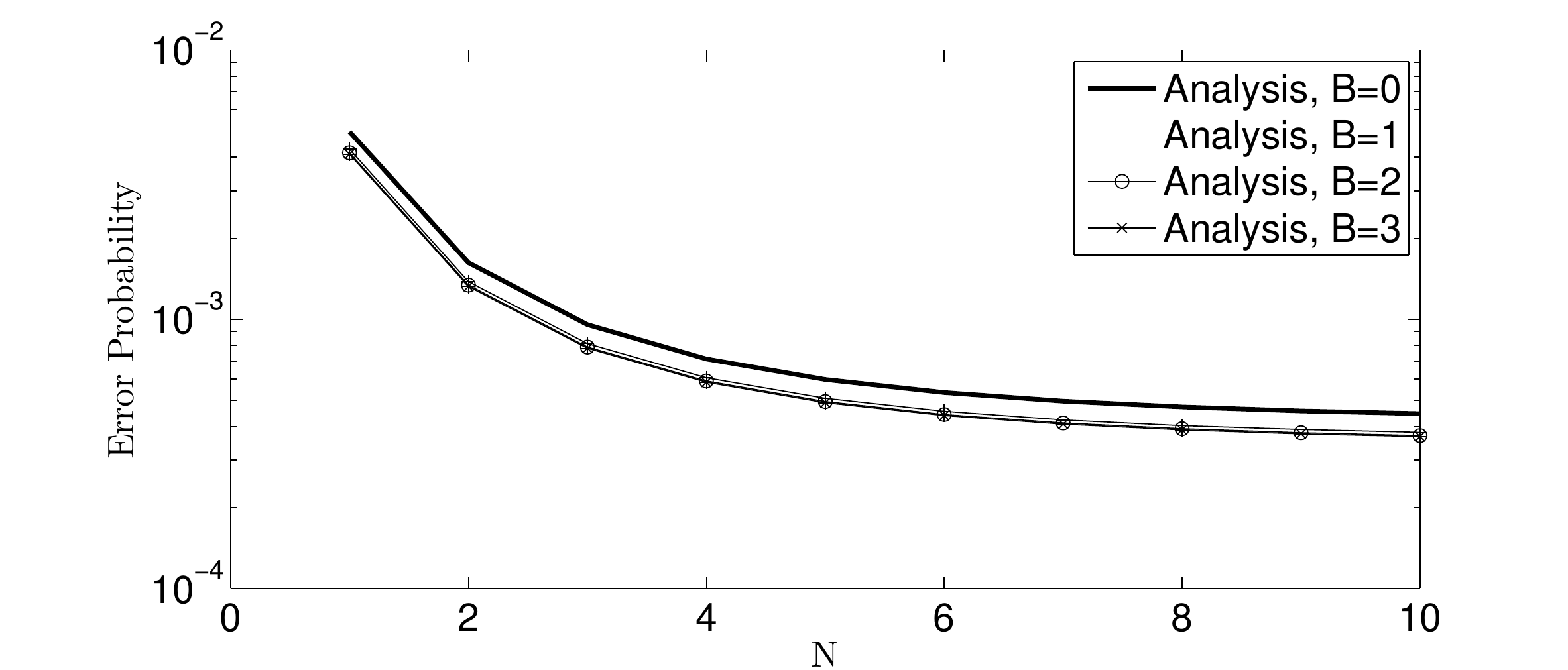}
\end{center}
\vspace{-1cm}
\caption{Probability of error for zero delay decoder with multiple reads in terms of $N$. Parameters for this figure: $s=40$ and $\lambda_0 = 4$, $t_s=5 sec$}
\label{figBN}
\end{figure}

\iffalse
\begin{figure}
\begin{center}
\includegraphics[width=190mm]{MYMYFIG1.pdf}%%%%%%%%%%%%%%%% 4
\end{center}
\caption{{Lower and upper bounds for limited memory receivers (for MCSK). The curves labeled as ``Threshold Dec." are lower bounds on $E_T(R, s, B)$. And the curve labeled as ``Known Int." is the lower bound on $E(R, s, B)$.  Parameters for this figure: $s=60$ and $\lambda_0=4$}}
\label{fig4}
\end{figure}
\fi

\begin{figure}
\begin{center}
\includegraphics[width=121mm]{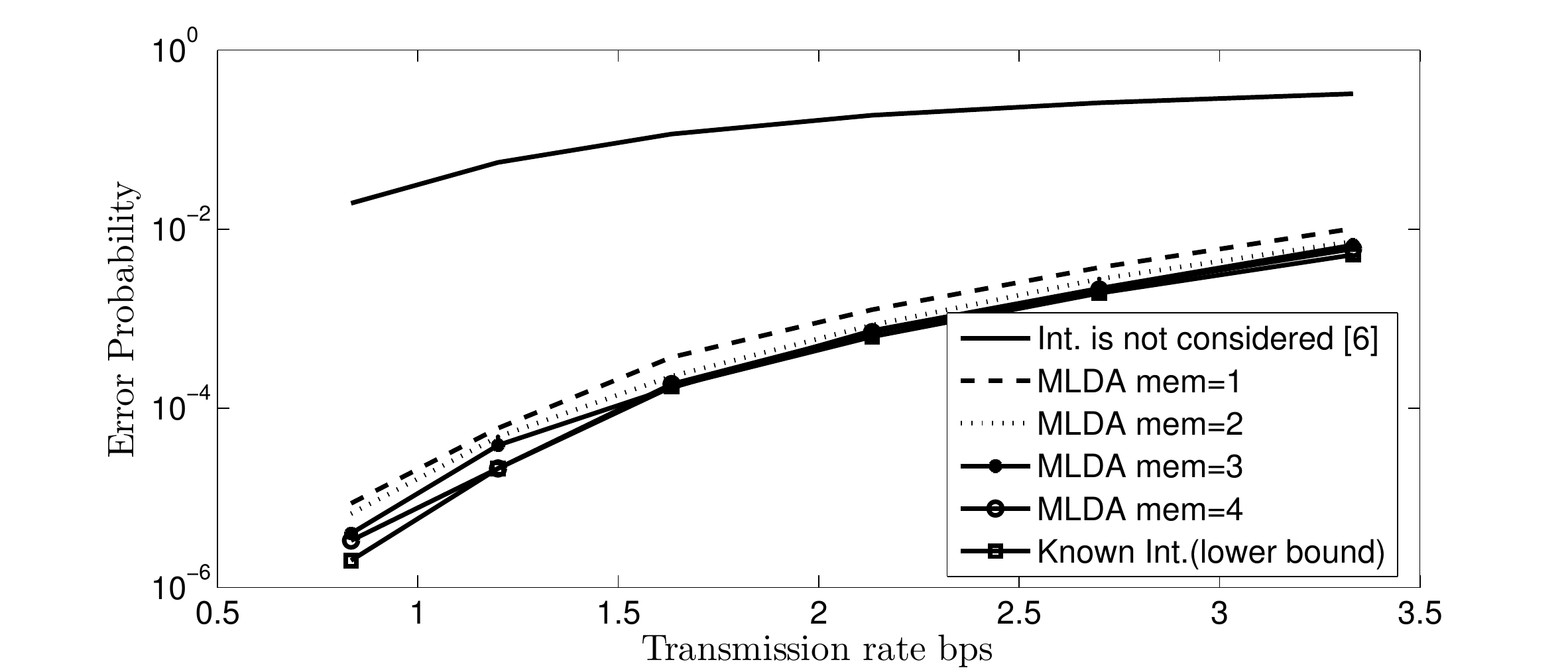}%%%%%%%%%%%%%%%% 4
\end{center}
\vspace{-1cm}
\caption{{Performance of MLDA for different memory size along with the lower bound (for MCSK). \iffalse The curve labeled by "Int. is not considered" is the performance of original MCSK modulation scheme proposed in \cite{AA1} where we do not consider the interference term in the calculations, and the curve labeled as ``Known Int." is the lower bound on $E(R, s, B)$. Parameters for this figure: $s=60$ and $\lambda_0=0.4$\fi} }
\label{fig4}
\end{figure}

\begin{figure}
\begin{center}
\includegraphics[width=135mm]{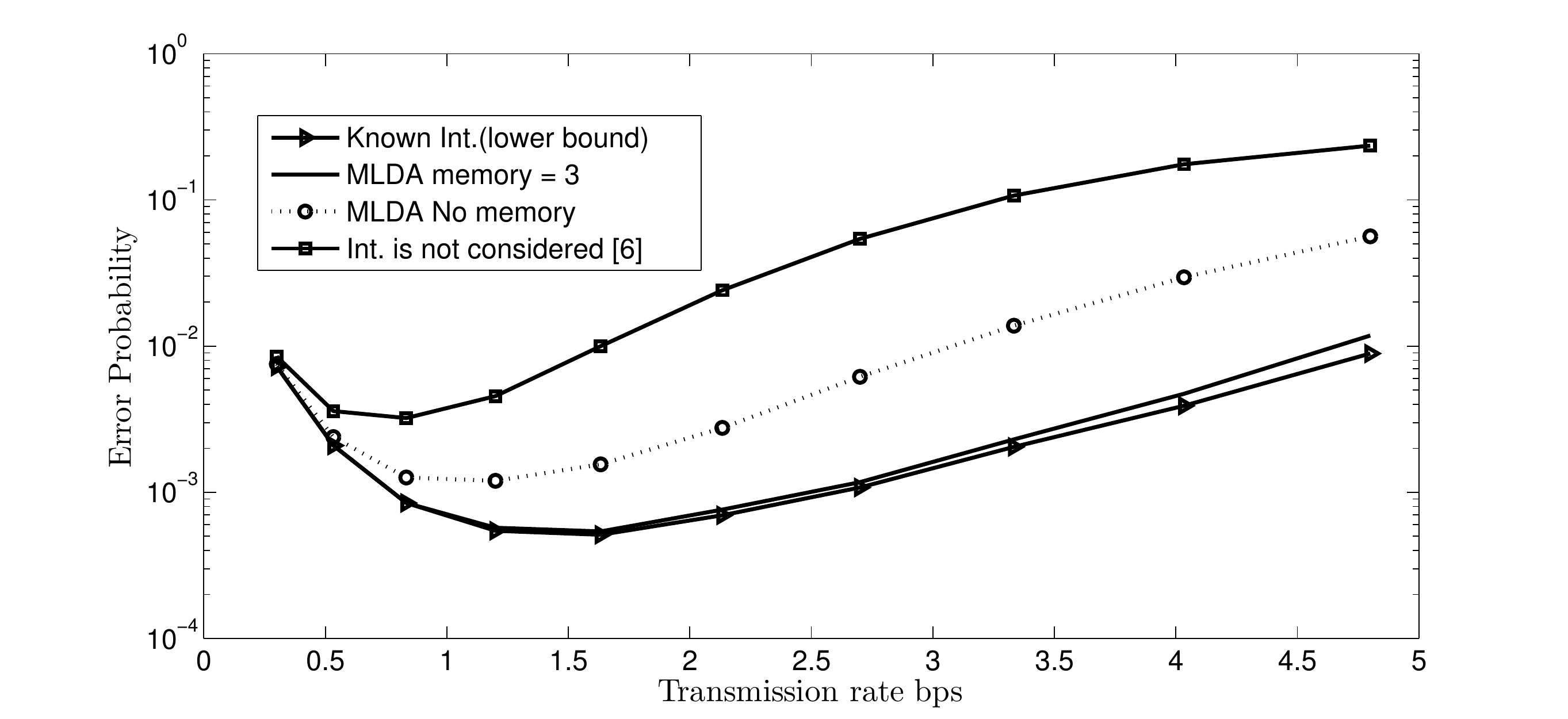} %%%%%%%%%%%%%%%%%%%% 5
\end{center}
\vspace{-1cm}
\caption{{Performance of MLDAs for large background noise. \iffalse The probability of error is not always monotone in transmission rate. Parameters for this figure:  $s=80$, $\lambda_0 =40$\fi }
}
\label{fig5}
\end{figure}

\iffalse
\begin{figure}
\begin{center}
\includegraphics[width=150mm]{All_2.pdf}            %%%%%%%%%%%%%%%%%%%% 7
\end{center}
\vspace{-1cm}
\caption{MLDA and ``Threshold Dec."  (lower bound on $E_T(R, s, B)$)  curves receivers for MCSK. Parameters for this figure: $s=60$, $\lambda_0 =4$} 
\label{fig7}
\end{figure}

\fi

\begin{figure}
\begin{center}
\includegraphics[width=93mm]{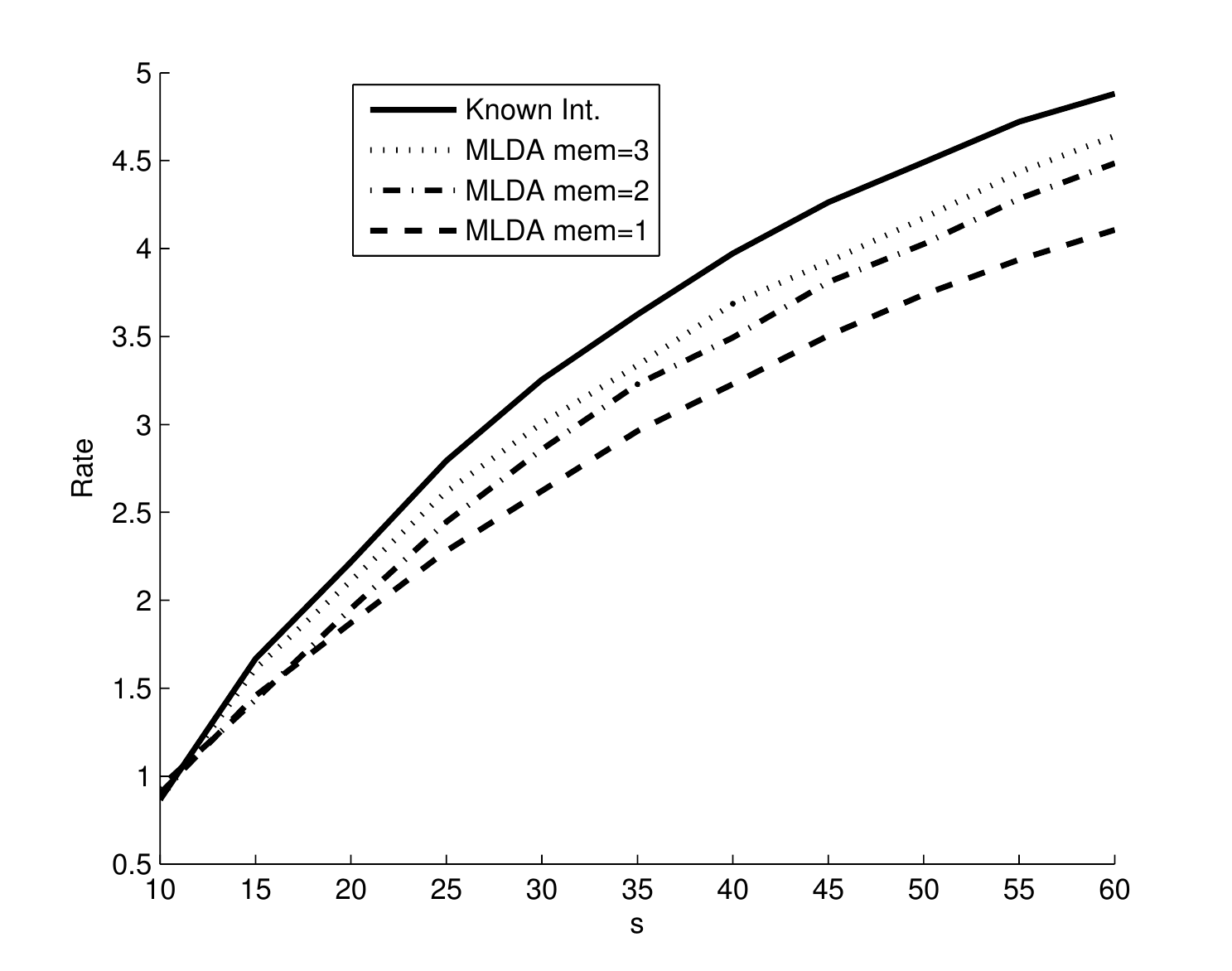}            %%%%%%%%%%%%%%%%%%%% 7
\end{center}
\vspace{-1cm}
\caption{
The maximum possible rate when the probability of error is less than or equal to 0.005 for a given $s$.\iffalse This is ploted for the known interference lower bound and MLDAs with different memories. Parameters for this figure:  $\lambda_0 \simeq 0$
new figure missing\color{black}\fi
} 

\label{fig7_mid}
\end{figure}

\begin{figure}
\begin{center}
\includegraphics[width=150mm]{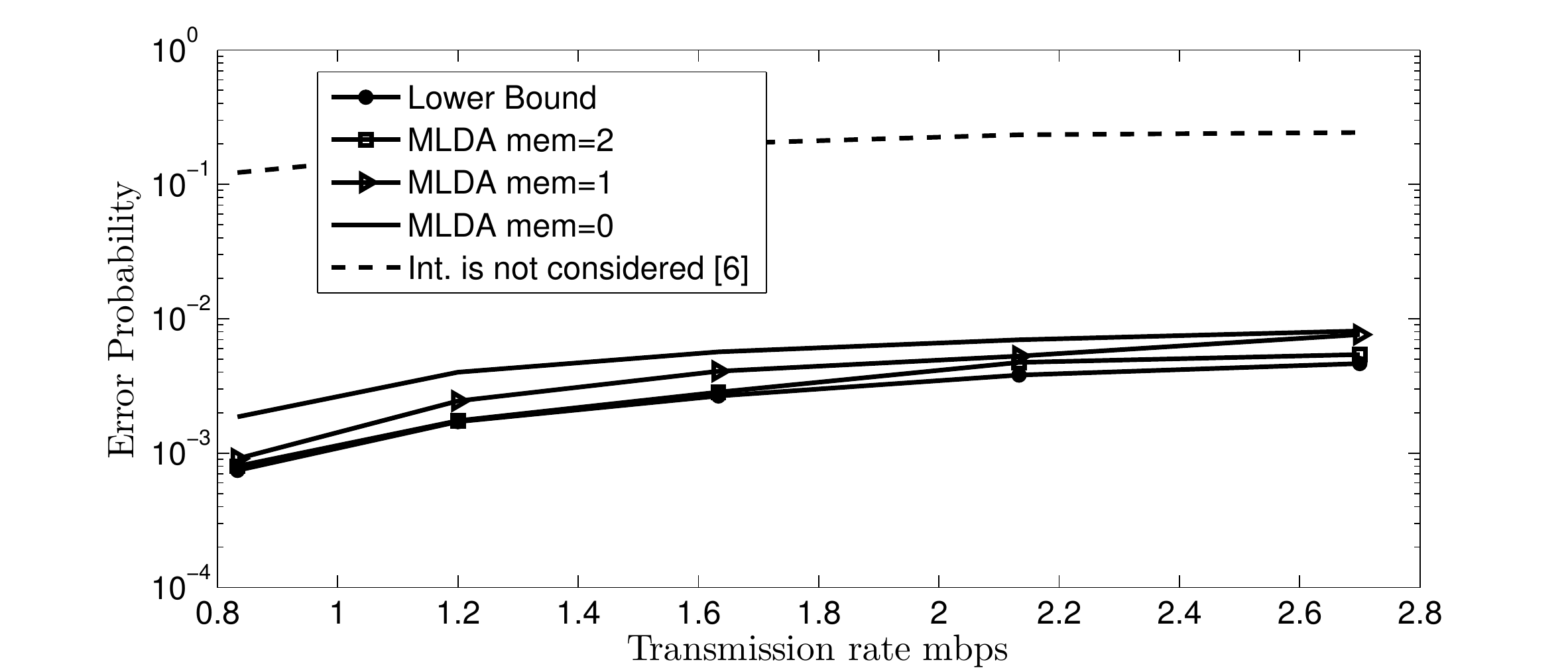}            %%%%%%%%%%%%%%%%%%%% 7
\end{center}
\vspace{-1cm}
\caption{{Performance of MLDA for different memory size along with the lower bound for 3-D medium (for MCSK). Parameters for this figure: $s=120$ and $\lambda_0 =0$} 
%new 3d figure \color{black}
} 

\label{Fig3d}
\end{figure}

\begin{figure}
\begin{center}
\includegraphics[width=155mm]{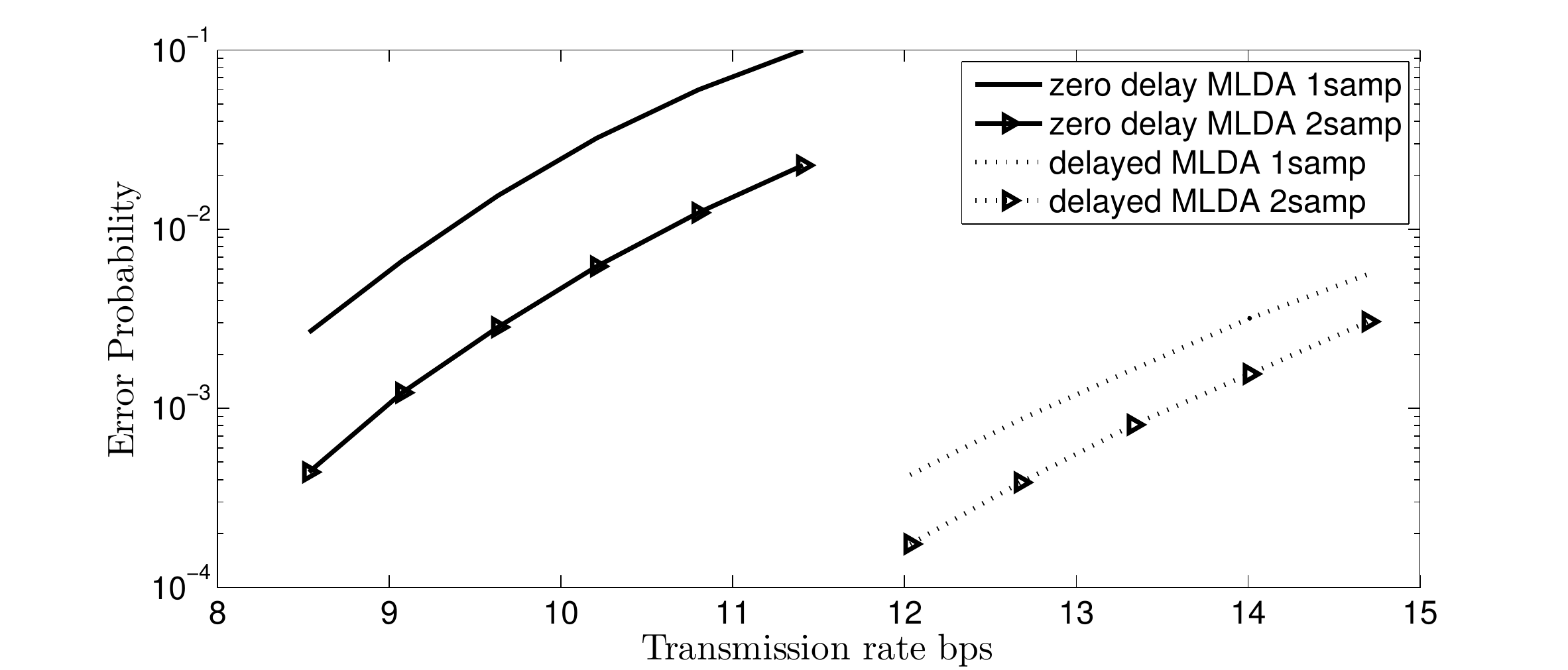}
\end{center}
\vspace{-1cm}
\caption{Probability of error for delayed and zero decoding with oversampling. Parameters for this figure: $B=4$, $s=800$ and $\lambda_0 = 40$ }
\label{fig12}
\end{figure}   %%%%%%%%%%%%%%%%%%%%%%%%%%%%%% 12

\iffalse
\begin{figure}
\begin{center}
\includegraphics[width=190mm]{3AllRates30_0_4.pdf} %%%%%%%%%%%%%%%%%%%%%%%%% 10
\end{center}
\caption{Probability of error of different proposed receivers in terms of transmission rate ( bps). Parameters for this figure:  $s=60$ and $\lambda_0=4$. }
\label{fig10}
\end{figure}

\begin{figure}
\begin{center}
\includegraphics[width=190mm]{mor3_fig5_2.pdf}
\end{center}
\caption{Probability of error of zero delayed and delayed decoding when the interference is considered and not considered. Parameters for this figure:  $B=4$, $\lambda_0 = 40$ and $t_s = 150ms$. }
\label{fig11}
\end{figure}   %%%%%%%%%%%%%%%%%%%%%%%%%%%%%% 11

\begin{figure}
\begin{center}
\includegraphics[width=190mm]{Samples_Pow400_Mem3.pdf} %%%%%%%%%%%%%%%%%%% 9
\end{center}
\vspace{-1cm}
\caption{Effect of multiple reads on reducing the probability of error of MLDA (for MCSK). Parameters for this figure: $s=800$ and $\lambda_0 = 40$}
\label{fig9}
\end{figure} %%%%%%%%%%%%%%%%%%%% 9

\begin{figure}
\begin{center}
\includegraphics[width=190mm]{figFNS.pdf}
\end{center}
\caption{Probability of error for zero delay decoder with multiple reads in terms of $s$. Parameters for this figure: $B=1$ and $\lambda_0 = 4$, $t_s=5 sec$}
\label{figNS}
\end{figure}
\fi

\vspace{-0.3cm}
\begin{thebibliography}{15}

\bibitem{AAA}I. F. Akyildiz, J. M. Jornet, and M. Pierobon, ``Nanonetworks: A new frontier in communications," \emph{ Communications of the ACM}, vol. 54, no. 11, pp. 84-89, November 2011.

\bibitem{Nakano}
T. Nakano, M.J. Moore,  F. Wei,  A.V.  Vasilakos, J. Shuai, ``Molecular Communication and
Networking: Opportunities and Challenges", IEEE Transactions on NanoBioscience, 11 (2), pp. 135-148, 2012.

\bibitem{B5}M. Pierobon and I. F. Akyildiz, ``A physical end-to-end model for
molecular communication in nanonetworks," \emph{IEEE Journal of Selected
Areas in Communications}, vol. 28, pp. 602–611, 2010.

\bibitem{New_2} A.W. Eckford, ``Nanoscale communication with Brownian motion", 41st Annual Conf. on Inf. Sci. and Sys., pp. 160-165.

\bibitem{AA2}M. S. Kuran, H. B. Yilmaz, T. Tugcu, I. F. Akyildiz, ``Modulation techniques for communication via diffusion in nanonetworks," \emph{Int. Conf. on Comm.}, June 2011.   

\bibitem{AA1}H. Arjmandi, A. Gohari, M. Nasiri-Kenari and Farshid Bateni, ``Diffusion based nanonetworking: A new modulation technique and performance analysis," \emph{IEEE Communications Letters}, 17(4):645 - 648, 2013.  

\bibitem{Eckford22} K. V. Srinivas, Raviraj S. Adve, and Andrew W. Eckford, ``Molecular Communication in Fluid Media: The Additive Inverse Gaussian Noise Channel," IEEE Trans. Inform. Theory,  58(7):4678-4692, 2012.

\bibitem{B1}M. S. Kuran, H. B. Yilmaz, T. Tugcu, B. Ozerman, ``Energy model for communication via diffusion in nanonetworks" 
\emph{Nano Communication Networks} 1(2):86-95, 2010.

\bibitem{B2}M. J. Moore, T. Suda, K. Oiwa, ``Molecular communication: Modeling noise effects on information rate", \emph{IEEE Trans. NanoBioscience}, 8(2):169-180, 2009.

\bibitem{AA3}H. Shahmohammadian, G. G. Messier, S. Magierowski, ``Optimum receiver for molecule shift keying modulation in diffusion-based molecular communication channel", \emph{Nano Communication Networks} 3:183-195, 2012.

\bibitem{NewlyAdded}A. Noel, K.  C. Cheung and R. Schober, ``Optimal Receiver Design for Diffusive Molecular Communication with Flow and Additive Noise", arXiv:1308.0109.

\bibitem{MY1} S. Kadloor, R. S. Adve, and A. W. Eckford, ``Molecular communication using brownian motion with drift," arXiv:1006.3959.

\bibitem{B3}M. S. Kuran, H.B. Yilmaz, T. Tugca, I. F. Akyildiz, ``Interference effects on modulation techniques in diffusion based networks", \emph{Nano Comm. Net.} 3(1):65-73, 2012. 

\bibitem{Book}
T. Nakano, A. W. Eckford, and T. Haraguchi, ``Molecular Communication", Cambridge University Press,  2013. 


%\bibitem{B4}M. S. Kuran, H. B. Yilmaz, T. Tugcu, I. F. Akyildiz, ``Modulation techniques for communication via diffusion in nanonetworks", 2011 IEEE International Conference on Communications (ICC), 2011, pp. 1-5.

\bibitem{B6}M. Pierobon and I. F. Akyildiz, ``Diffusion-based noise analysis for molecular communication in nanonetworks", \emph{IEEE Transaction on Signal Processing}", 59(6):2532-2547, 2011.


%\bibitem{New} R. S. Chhikara and J. L. Folks, ``The Inverse Gaussian Distribution: Theory, Methodology, and Applications." New York: Marcel Dekker, 1989.

\bibitem{saddle} C. W. Helstrom ``Approximation evaluation of detection probabilities in radar and optical communications," \textit{IEEE Trans. of Aerospace and Electronic Systems}, 14(4): 630-640, 1978.

\bibitem{Web}A. Crofts, ``Diffusion", available at http://www.life.illinois.edu/crofts/bioph354/diffusion1.html

\vspace{-0.6cm}
 {\bibitem{Re1}P.-C. Yeh, K.-C. Chen, Y.-C. Lee, L.-S. Meng, P.-J. Shih, P.-Y. Ko, W.-A. Lin and C.-H. Lee, ``A new frontier of wireless communication theory: diffusion-based molecular communications."\emph{IEEE Wireless Communications}, 19 (5): 28 - 35 , 2012.}

\vspace{-0.7cm}
 {\bibitem{Isomer}N.-R. Kim, C.-B. Chae, ``Novel modulation techniques using isomers as messenger molecules for nano communication networks,"  \emph{IEEE JSAC}", 31 (12):847 - 856 , 2013.}
\end{thebibliography}
\end{document}